\documentclass[acmlarge]{acmart}
\usepackage{url}
\usepackage{booktabs} %
\usepackage{listings}
\usepackage{graphics}   
\usepackage{comment}
\usepackage{colortbl}
\usepackage[final]{changes}
\usepackage{subcaption}
\usepackage{caption} 
\usepackage{algpseudocode}
\usepackage[ruled]{algorithm2e} %
\newcommand{\step}[1]{{\color{ACMRed}#1}}

\algnewcommand\algorithmicinput{\textbf{Input:}}
\algnewcommand\INPUT{\item[\algorithmicinput]}

\algnewcommand\algorithmicoutput{\textbf{Output:}}
\algnewcommand\OUTPUT{\item[\algorithmicoutput]}

\newcommand{\modelname}[1]{{\it ``#1''}\xspace}

\SetAlFnt{\small}
\SetAlCapFnt{\small}
\SetAlCapNameFnt{\small}
\SetAlCapHSkip{0pt}
\IncMargin{-\parindent}

\acmJournal{IMWUT}
\acmYear{2019} 
\acmVolume{3} 
\acmNumber{1} 
\acmArticle{34} 
\acmMonth{3} 
\acmPrice{15.00}
\acmDOI{10.1145/3314421}

\setcopyright{acmcopyright}
\received{August 2018}
\received{November 2018}
\received[accepted]{January 2019}

\graphicspath{ {graphics/} }

\newcommand{\etal}{{\em et al. }} 

\newcommand{\systemname}{GEVR}

\begin{document}

\title[\systemname: An Event Venue Recommendation System for Groups of Mobile Users]{\systemname: An Event Venue Recommendation System for Groups of Mobile Users} 

\author{Jason Shuo Zhang}
\email{jasonzhang@colorado.edu}
\orcid{1234-5678-9012-3456}
\affiliation{%
  \institution{University of Colorado Boulder}
  \streetaddress{1111 Engineering Dr}
  \city{Boulder}
  \state{CO}
  \postcode{80309}
  \country{USA}
}

\author{Mike Gartrell}
\email{Mike.Gartrell@colorado.edu}
\affiliation{%
  \institution{University of Colorado Boulder}
  \streetaddress{1111 Engineering Dr}
  \city{Boulder}
  \state{CO}
  \postcode{80309}
  \country{USA}
}

\author{Richard Han}
\email{richard.han@colorado.edu}
\affiliation{%
  \institution{University of Colorado Boulder}
  \streetaddress{1111 Engineering Dr}
  \city{Boulder}
  \state{CO}
  \postcode{80309}
  \country{USA}
}

\author{Qin Lv}
\email{qin.lv@colorado.edu}
\affiliation{%
  \institution{University of Colorado Boulder}
  \streetaddress{1111 Engineering Dr}
  \city{Boulder}
  \state{CO}
  \postcode{80309}
  \country{USA}
}

\author{Shivakant Mishra}
\email{shivakaht.mishra@colorado.edu}
\affiliation{%
  \institution{University of Colorado Boulder}
  \streetaddress{1111 Engineering Dr}
  \city{Boulder}
  \state{CO}
  \postcode{80309}
  \country{USA}
}

\begin{abstract}
In this paper, we present \systemname{}, 
the first Group Event Venue Recommendation system 
that incorporates mobility via individual location traces 
and context information into a \modelname{social-based} group decision model to provide venue recommendations
for groups of mobile users.
Our study leverages a real-world dataset collected using the OutWithFriendz mobile app for group event planning, which contains 625 users and over 500 group events. 
We first develop a novel \modelname{social-based} group location prediction model,
which adaptively applies different group decision strategies 
to groups with different social relationship strength to aggregate each group member's
location preference, to predict where groups will meet.
Evaluation results show that our prediction model not only outperforms commonly used and state-of-the-art group decision strategies with over 80\% accuracy for predicting groups' final meeting location clusters, 
but also provides promising qualities in cold-start scenarios. 
We then integrate our prediction model with the Foursquare Venue Recommendation API 
to construct an event venue recommendation framework for groups of mobile users. 
Evaluation results show that \systemname{} outperforms the comparative models by a significant margin. 
\end{abstract}

\ccsdesc[500]{Human-centered computing~Ubiquitous and mobile computing}
\keywords{Group Recommendation, Event Venue, Group Location Cluster, User Mobility}

\maketitle

\renewcommand{\shortauthors}{J.S. Zhang et al.}
\section{Introduction}
\label{sec:intro} 
Online venue recommendation services, such as Foursquare, Google Maps, and Yelp, demonstrate the importance placed on recommending venues to an individual mobile user, e.g., what coffee shop or restaurant should that person visit next.  However, such venue recommendation does not address the larger challenge of recommending venues to a group of mobile users who are trying to choose a place to meet, e.g., friends and colleagues trying to decide on a restaurant for lunch or dinner.  In this case, the members of the group may have a wide range of behaviors and preferences and be scattered across many locations.

This paper addresses this topic of recommending event venues to {\it groups of mobile users}, a service that we expect will become increasingly important as people get more and more digitally connected and organize their social gatherings via mobile technologies. While effective recommendation to individual users is already a challenging task, developing event venue recommendations for groups of mobile users is significantly more challenging due to a number of reasons.  Groups of users often have a wide variety of behaviors and preferences which must be resolved to achieve the best recommendations to the group.  In addition, different members of a group may be at varied locations, so mobility considerations should be taken into account when recommending a place to meet.  Groups are also dynamic rather than static, and so group membership may change fluidly from event to event, unlike individual recommendation for the same target user.

Prior studies in group-based recommendation has addressed the issue of resolving differing preferences of individual members of the group in order to provide the best recommendation~\cite{gartrell2010enhancing,chaney2014large,quintarelli2016recommending}. However, the role that location and mobility of different group members may play in influencing the group's final choice has not been addressed in these articles. Some scant prior research in group event venue recommendation~\cite{beckmann2011agremo,park2008restaurant} has been confined only to in-lab surveys 
and has not studied the real-world dynamics of groups of mobile users, 
nor its impacts on group decision-making.   We believe this is an exciting area ripe for exploration by the ubiquitous computing research community.  Due to the explosion of people scheduling group events online using smart devices, 
we expect mobile and context-aware computing to be used extensively to reduce the friction inherent in coordination and assist groups in making decisions.

As the first step towards this direction, we present \systemname{}, 
a novel Group Event Venue Recommendation system that
tackles the recommendation challenge by incorporating mobility patterns 
and social relations among group members. 
To tackle the complexity of group recommendation, 
\systemname{} splits the recommendation process into three separate steps:
\begin{enumerate}
\item {\bf Group location cluster detection:} Using individual group members' historical location traces,
\systemname{} first identifies location clusters where the group members have colocated in the past. 
\item {\bf Group event location cluster prediction:} 
After identifying these group location clusters, \systemname{} uses
a newly designed social-based group location cluster prediction model to estimate 
which location cluster the group is more likely to meet at.
\item {\bf Event venue recommendation:} Finally,
\systemname{} aggregates restaurants near group members' frequented locations 
using the Foursquare Venue Recommendation API~\cite{foursquare},
and re-ranks them based on the prediction results calculated by Step (2) 
to build a final top-K recommendation list.
\end{enumerate} 
Evaluation results show that our prediction model achieves over 80\% accuracy for predicting a final meeting location cluster. 
Furthermore, through the combination of the group location clusters identified, cluster prediction results and Foursquare Venue Recommendation API,
\systemname{} achieves promising recommendation performance and 
outperforms the baseline models by a significant margin.
A key contribution of this paper is that the efficacy of \systemname{} has been evaluated using a large-scale, real-world dataset that was
collected from a mobile application called OutWithFriendz~\cite{zhang2018understanding}, which supports group event scheduling by enabling users to propose event venues and dates, invite people to join the group, and vote on venues and dates.  The finalized event location and date are confirmed through location tracking in the app.

In summary, we make the following fundamental contributions in this work:
\begin{itemize}
\item We introduce GEVR, the first system that provides event venue recommendation to groups of mobile users. We present a novel algorithm for recommending event venues to groups of mobile users that re-prioritizes the results of the Foursquare Venue Recommendation API according to group properties, namely the likelihood of the venue residing in a location \emph{cluster} that the group is likely to visit.  
\begin{itemize}
\item To generate a ranking of each group's preferences for its location clusters, we separate the problem into first generating individual group members' preferences for those clusters, and then combining the individual preferences into a group ranking via a group decision strategy.
\item Our analysis reveals that location familiarity, individual activeness, day of the week, 
and population density are strong predictors of individual cluster preferences.
\item We construct an innovative \modelname{Edge-RWR+CART} model to incorporate these identified features
for predicting individual location preference. 
It is not only demonstrated to be more superior to the comparative models 
but also flexible for adding new contextual features.
\item We design a novel \modelname{social-based} group location prediction model, which adaptively applies different group decision strategies to groups with different social relationship strength to aggregate group members' individual location preferences. 
\end{itemize}
\item We collect data that is difficult to attain concerning which venues groups choose to meet at, 
obtaining over 500 real-world group events by over 600 users from a mobile application over one year, 
thereby establishing ground truth on which venues these groups chose for their gatherings.
\item We evaluate the effectiveness of GEVR using our real-world dataset. 
Detailed evaluation results show that using our \modelname{social-based} group location prediction model 
outperforms all the commonly used and state-of-the-art baselines with a significant margin, 
even in the cold-start scenarios.

\end{itemize}

The remainder of this paper is organized as follows. In the next section, we discuss related work. 
In Section~\ref{sec:data}, we review the functions of the group event mobile application relevant to our 
dataset collection, and the problem formulation. 
Then in Section~\ref{sec:cluster}, we present detailed analysis of group location cluster detection 
and parameter selection using collected user location traces. 
The design of our social-based group location prediction model is introduced in Section~\ref{sec:prediction}, 
and the framework of our group event venue recommendation is described in Section~\ref{sec:recommender}. 
Performance evaluations are presented in Section~\ref{sec:evaluation}. 
Finally, we highlight the key findings, discuss important implications and future improvements, and conclude this paper.

\section{Related Work}
\label{sec:related}

The problem of making recommendations to a group of users has been investigated in a number of 
works~\cite{gartrell2010enhancing,chaney2014large,quintarelli2016recommending}. Gartrell \etal
propose a group recommendation model that utilizes social and content interests of group members
to recommend movies for groups of users~\cite{gartrell2010enhancing}. The work by Cheney \etal
presents a large-scale study of television viewing habits. 
They provide an analysis of how the viewing
patterns shift across various group contexts and discussed the impact of these findings on the performance
of a group TV program recommendation system~\cite{chaney2014large}. Quintarelli \etal further
models users' ability
to direct a group's decision by using contextual influence and aggregate it to recommend TV programs
for groups of users~\cite{quintarelli2016recommending}. All of these works can be summarized as 
recommending items for groups of users, where user mobility is not much of a concern. 
Our problem, group event venue recommendation is more challenging
as users' mobility patterns can play a significant role in group decision making. 
Before the actual event gathering, 
group members may be distributed in different geographical areas. 
Thus, recommending a venue that will 
optimize group satisfaction is more challenging and requires delicate data analysis and system design.

Our work is also generally related to the research topic of understanding the relationship between online
and offline social behaviors in groups. 
Lane~\etal consider various social phenomena
and environmental factors and design a Networked Community Behavior framework which
can exploit community-scale behavior 
patterns~\cite{lane2014connecting}. 
Zhang~\etal studies how does team performance impact fan 
behavior on Reddit NBA fan groups~\cite{zhang2018we}.
The work by Park~\etal introduce a social context-aware smart-phone notification system to help groups
of users focus better on in-person social interactions~\cite{park2017don}. Jayarajah \etal 
investigate how users deal with group co-presence to prevent conflictive 
situations on campus~\cite{jayarajah2015need}. 
There are also a few works studying Doodle~\cite{reinecke2013doodle, zou2015strategic,romero2017influence}, 
an online event scheduling service, 
where the host can only suggest the me{}eting times for group members to vote, not venues.
Our work adds to this literature by demonstrating the effects of different key factors on group 
event attendance and how to aggregate individual location preferences to model group decision-making.
 
Earlier work in event organization has investigated a series of contextual features 
that would impact a user's decision to attend a group event 
~\cite{macedo2015context,du2014predicting,georgiev2014call,Zhang:2019:UEO:3306498.3243227,zhang2018hybrid}. 
Most of these studies rely on datasets collected from 
popular group event services, 
such as Meetup,~\footnote{https://www.meetup.com/} Evite,~\footnote{https://www.evite.com/} 
and Douban Events.~\footnote{https://beijing.douban.com/events/} 
This line of research still falls into individual recommendation 
as the major task is to match potential users who are interested in the event.
Very little existing research has explored the subject of recommending event venues for groups of mobile users. 

Finally, the proliferation of location-based social services in recent years has enabled a rich body of research 
in understanding individual movement patterns and predicting when and where people will go next, such as next place prediction~\cite{baumann2013influence,noulas2012mining},
trip planning~\cite{zheng2009mining,lu2010photo2trip},
ride-sharing management~\cite{cici2014assessing,agatz2012optimization,chen2016dynamic},
and individual venue recommendation~\cite{noulas2012random,lian2014geomf}.

\section{Dataset}
\label{sec:data}

The dataset we use in this work is collected using the OutWithFriendz mobile application, which enables group hosts to organize events and negotiate when and where to meet with group members through an embedded voting process~\cite{zhang2018understanding}.
The app is implemented as a client-server architecture that is comprised of both Android and iOS based clients.
When creating an event, the host can specify the details of the invitation,
including a title, list of suggested meeting venues, times, and invited participants.
After the host creates and submits a new invitation, all participants
will receive a notification. Then the event scheduling process begins.
Every invitee can vote for their preferred venues and times, and suggest more options.
After the group consensus process ends, the host decides
the final meeting venue and time. To support later negotiation,
the host can still update the final venue and time after it is finalized.
In the design of the OutWithFriendz mobile application,
the host can make decisions based on
the voting results, but he/she does not always have to obey them.

The whole dataset contains group event data collected from January 2017 to December 2017. 
We use two methods for data collection: 
(1) We advertise the mobile application in our university's student center, dining rooms,
and departmental lobbies.
(2) We post task assignments on two crowdsourcing platforms:
Microworkers~\cite{microworkers} and Craigslist~\cite{craigslist}. 
For the users of crowdsourcing platforms, \$10--20 is paid to the
group host for completing a legitimate event using OutWithFriendz,
with the provisions that:
\begin{enumerate}
\item The host must live in the US.
\item The host should invite at least two other members to the invitation using the mobile app. 
\item The meeting venue and time must be finalized. 
\item Each group member should open the location service on his/her smart phone 
during the event scheduling process and allow location traces to be collected by the mobile app.
\item At least half of the group members should attend the finalized event. 
We manually confirm this using location traces 
before approving workers' assignments on crowdsourcing platforms.
\end{enumerate}
A user may create or join multiple invitations during this study. 
To collect user mobility data, 
the OutWithFriendz mobile app posts users' GPS location traces to the server every five minutes 
when the mobile app is running in the background 
or every 30 seconds when the app is running in the foreground.
The experimental protocol is reviewed and conducted under IRB Protocol \#12-0008.

\begin{table}[]
\centering
\caption{Dataset Statistics. The users in our dataset are from two sources. 
Campus users are advertised from our university and volunteered to use the mobile app. 
Crowdsourcing users are recruited from Microworkers and Craigslist. 
Our users are all over the US, across 40 states and 117 cities.}
\begin{tabular}{lrrr}
\toprule
         & Total   & Campus & Crowdsourcing \\ 
\midrule
Number of users    & 625  & 75     & 550              \\ 
Number of events & 502 & 151  & 351            \\
\bottomrule
\end{tabular}
\label{tab:stats}
\end{table}

\begin{figure}
\centering
\begin{subfigure}{0.35\textwidth}
  \centering
  \includegraphics[width=1\linewidth]{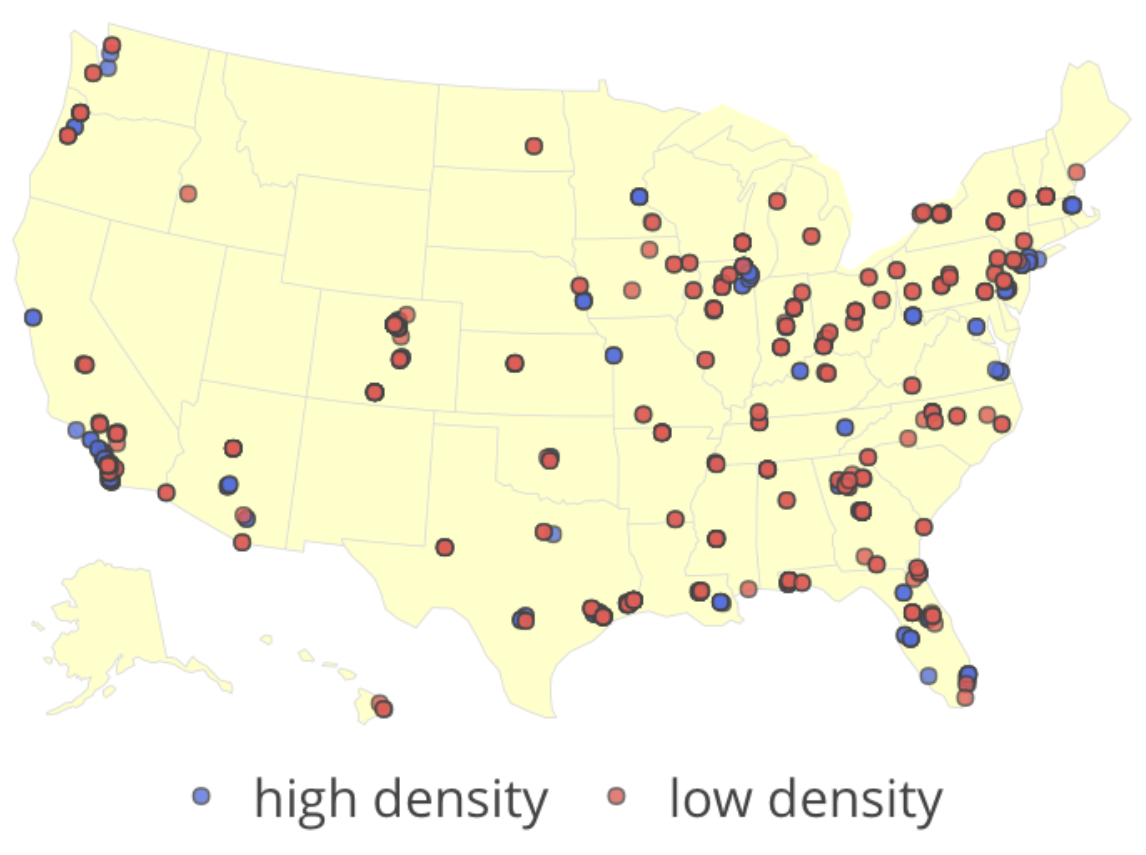}
  \caption{All the suggested venues across the US.}
  \label{fig:map}
\end{subfigure}%
\begin{subfigure}{0.65\textwidth}
  \centering
  \includegraphics[width=0.9\linewidth]{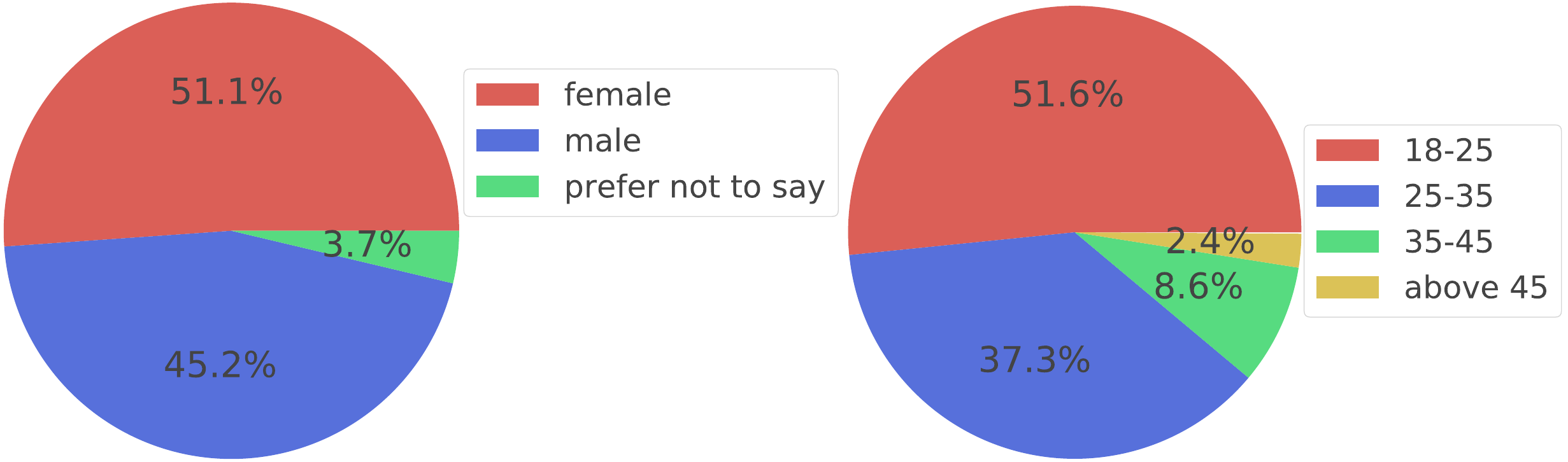} 
  \vspace{4em}
  \caption{User demographics.} 
  \label{fig:demo}
\end{subfigure}
\caption{Figure~\ref{fig:map} shows the geographic distribution of all the suggested venues across the US. 
Blue marks represent places of high-density areas and red in low-density areas. 
Figure~\ref{fig:demo} shows the demographics (gender and age) of 341 users in the dataset
who fill out the anonymized user demographic questionnaire. }
\label{fig:data}
\end{figure}

Since our~\systemname{} system utilizes users' location traces to provide recommendation for groups
of users, to ensure sufficient data coverage, we remove group events that have members with 
less than three days of location trace data before the invitation was created. 
Also, we remove group events in which the final meeting
venue is not a restaurant.~\footnote{Groups can organize any types of events using OutWithFriendz,
including going to the gym, watching a movie at the theater, or participating in outdoor activities. 
In this paper, we focus on dining events, as it is the dominating type in the dataset: 81\% of the events
are dining events.} 
The basic statistics of our dataset after the filtering are shown in Table~\ref{tab:stats}. 
The geographic distribution of all suggested venues across the US is projected in Figure~\ref{fig:map}.
All the venues are further divided into low density and high density areas based on the population
density of the venue location. Here the population density information is collected 
from the 2016 US area development degree data from the US Census Bureau~\cite{uscensus}. 
Our users are widespread over the US, covering 40 states and 117 cities.
After the user study, we conduct an anonymized user demographic questionnaire 
and 341 study participants completed it. 
Figure~\ref{fig:demo} shows these users' demographic information.
We also observe that all of the groups in our study are of either small or medium sized, 
with group sizes of 3--6. 
Among them, 64\% of the groups have three members, 19\% of the groups have four members,
12\% of the groups have five members, and the remaining 5\% have six members.

\section{Framework Overview}
\label{sec:frameoverview}
Given all the information collected from the mobile app, to
build an effective event venue recommendation system for groups,
we tackle the challenges in three steps:

\smallskip\noindent{\bf Step 1: Group location cluster detection.}
In this step, our objective is to gather location trace points of group members into location clusters. 
The detected location clusters should be able to represent group members' frequented locations.
The details of group location cluster detection will be introduced in Section~\ref{sec:cluster}.
The venues are searched and selected using the integrated Google Maps API.

\smallskip\noindent{\bf Step 2: Group location cluster prediction.}
Using the group location clusters detected from Step~\step{1}, 
our goal of this step is to predict which location clusters the group is likely to meet at. 
More specifically, we model the probability of each detected location cluster being selected 
as the final meeting location cluster of the group. 
The design of our prediction model will be explained in Section~\ref{sec:prediction}.

\smallskip\noindent{\bf Step 3: Event venue recommendation for groups of mobile users.}
In this step, our goal is to recommend a list of restaurants for each group event.
We use the Foursquare Venue Recommendation API to create a candidate restaurant pool 
near every detected location cluster generated by Step~\step{1}, 
and re-rank the candidate pool using the prediction results calculated
by Step~\step{2} to build a final top-N restaurant recommendation list.
The construction of our group event venue recommendation framework 
will be described in Section~\ref{sec:recommender}.

\smallskip To make effective event venue recommendation for 
groups of mobile users, the ultimate goal is to reduce and prioritize the venue
recommendation list. 
The three steps gradually build towards this goal.
We first use Step~\step{1} to detect group
members' frequented geographic areas and concentrate our recommendation
on these areas instead of the whole city. 
After that, by modeling group members' location preferences and aggregating them
using the group decision strategy in Step~\step{2}, we estimate
the likelihood of each location cluster that the group will decide to meet at.
Finally, in Step~\step{3}, we re-rank the candidate venues in concentrated
areas by the estimated likelihood of each location cluster.

\section{Group Location Cluster Analysis}
\label{sec:cluster}

Given a group with multiple mobile users and their location traces, 
our first goal is to identify {\em group location clusters}, i.e., concentrated regions that 
group members have visited.  Intuitively, a location cluster may represent home, work, 
or a district, say downtown, near a mall, or a zoned business area where there are many 
restaurants.  Due to the fact that user location data recorded 
by mobile applications usually contains errors of a few feet, 
it is impractical to use GPS points in location traces directly for modeling. 
A common approach in user mobility studies is 
to cluster nearby geolocation points into location 
clusters~\cite{ashbrook2003using, hsieh2013lifestreams,chen2016dynamic} 
and use them instead of location points when predicting human movements. 
Inspired by these papers, to generate group location clusters, 
we first combine all group members' temporally ordered location trace points and apply the widely-used DBSCAN~\cite{ester1996density} location clustering algorithm to detect group location clusters, 
which represent significant places that the group members visit frequently.
More specifically, \added{every location point is represented by 
its geographical coordinate pair (latitude and longitude), and fed directly into the DBSCAN clustering
algorithm, which} identifies dense neighborhoods and groups location points that are very close to each other into the same cluster to represent a location. 
After the clustering procedure, 
the ideal situation is that every location point detected 
from group members' location traces is assigned to a specific location cluster;  
nearby location points would belong to the same location cluster, 
and each cluster is relatively separated from each other on the map.
We further represent each group location cluster using the centroid
 of all the location trace points that belong to that cluster.
 To ensure that the DBSCAN algorithm is able to find sensible 
and meaningful location clusters using our dataset, and
the group's final meeting venue falls within one of these clusters, 
it is important to analyze and select appropriate values for the parameters. 

\begin{figure}
\centering
\begin{subfigure}{0.5\textwidth}
  \centering
  \includegraphics[width=1\linewidth]{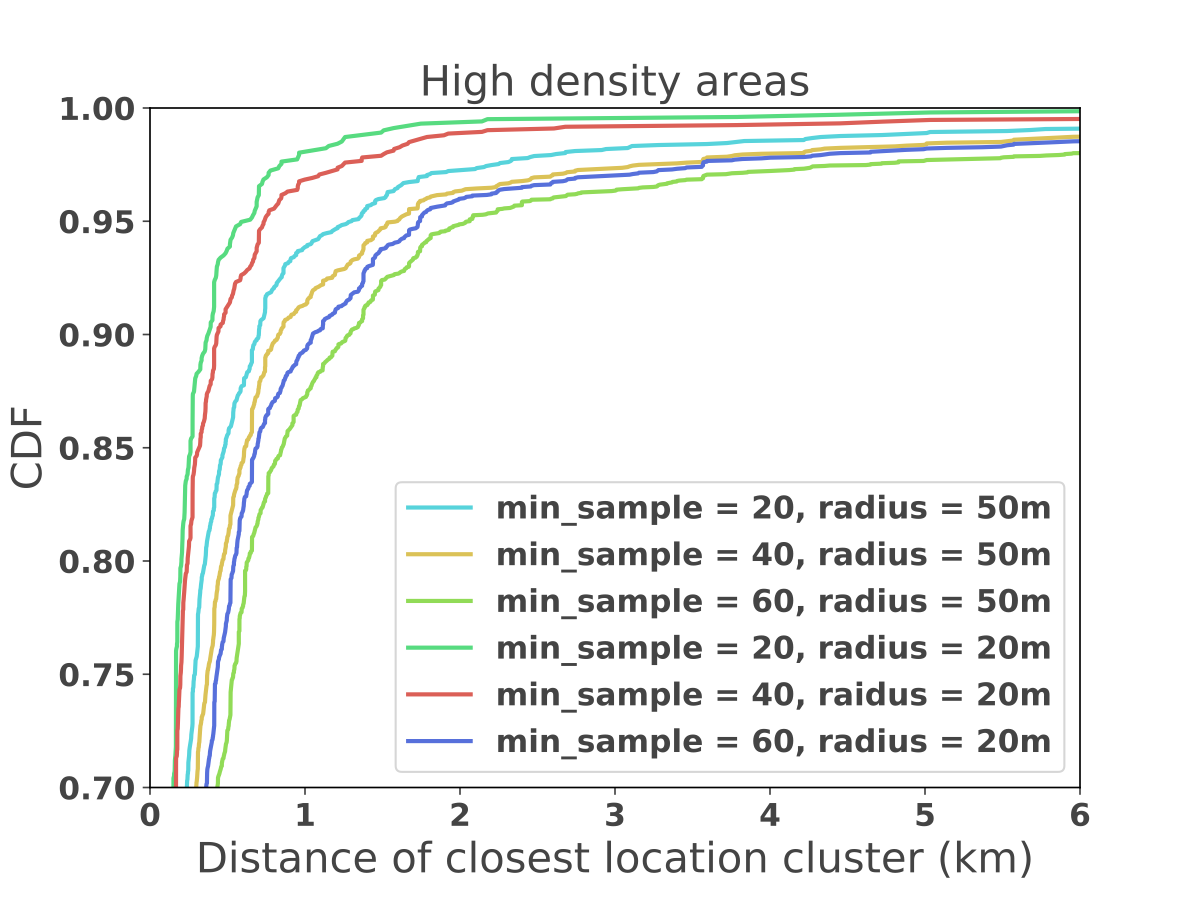}
  \caption{Distribution of distances in high density areas.}
  \label{fig:clusterhigh}
\end{subfigure}%
\begin{subfigure}{0.5\textwidth}
  \centering
  \includegraphics[width=1\linewidth]{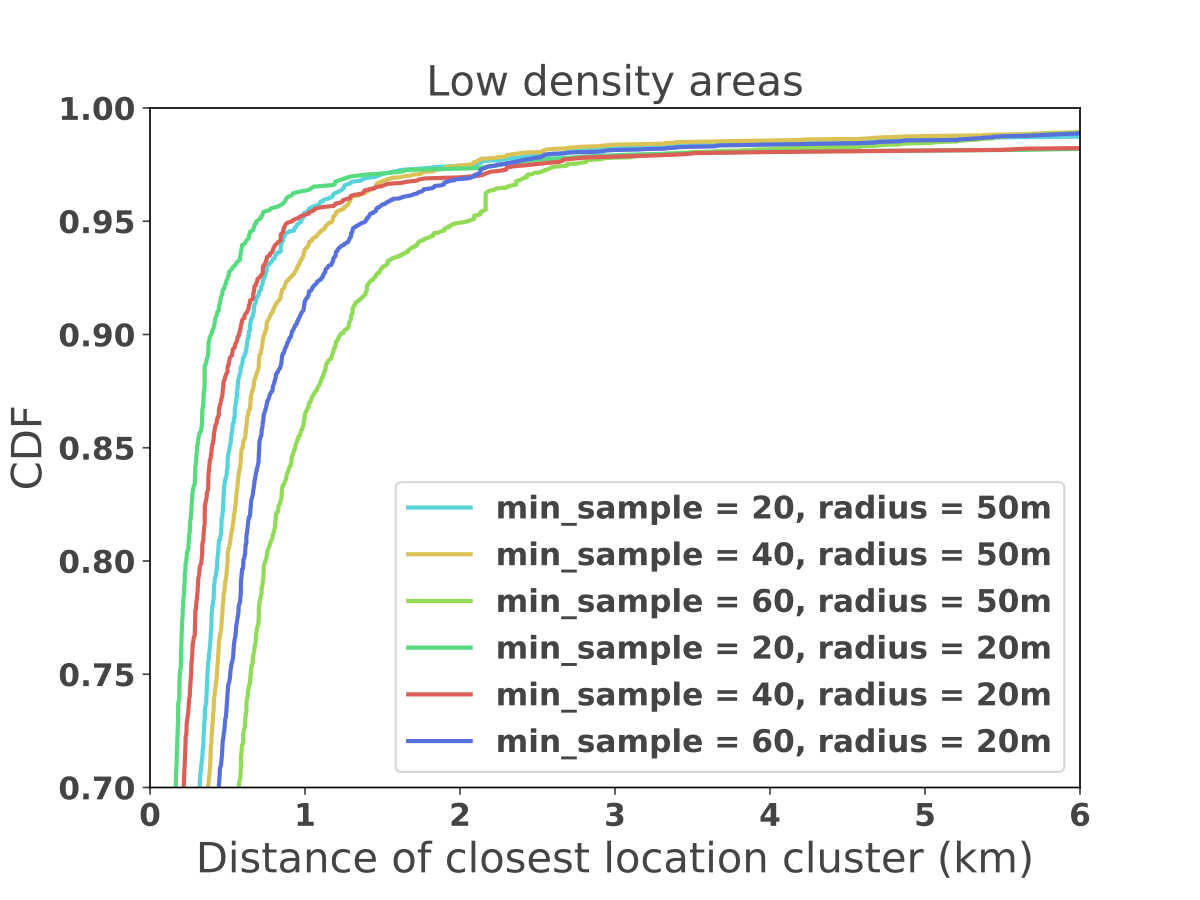} 
  \caption{Distribution of distances in low density areas.}
  \label{fig:clusterlow}
\end{subfigure}
\caption{Figure~\ref{fig:clusterhigh} and Figure~\ref{fig:clusterlow} 
show the cumulative distribution function of distances 
between each group event's final venue and the closest group location cluster 
for groups in high density and low density areas,  respectively. 
As can been seen in the figures, when setting $min\_sample=40, radius =\ 20 meters$,  
more than 90\% of group events' final venues 
fall within the closest group location cluster (within 500 meters).}
\label{fig:closest}
\end{figure}

The main parameters of DBSCAN are $min\_sample$ and $radius$. 
$Min\_sample$ is the minimum number of points required to form a dense region, and  
$radius$ is the maximum distance allowed between two samples for them to be considered 
as in the same neighborhood.
To determine a reasonable combination of $min\_sample$ and $radius$ parameter values, 
we tried different combinations and examined 
whether groups' final event venues fall within one of the group location clusters detected. 
Each group location cluster is represented by the centroid point 
of all location trace points that belong to that cluster.
Figure~\ref{fig:closest} shows the cumulative distribution function (CDF) of distances 
between a group event's final venue and the closest group location cluster using different 
combinations of $min\_sample$ and $radius$ values. For comparison purpose,
all groups are also divided equally into low density and 
high density areas based on the population density of their centroid points. 
The population density information we use is collected from the 2016 US 
area development degree data from the US Census Bureau~\cite{uscensus}. 

We find that in both high density and low density areas, when setting 
$min\_sample=40, radius=20\ meters$,  more than 90\% of the final event venues 
fall into the closest clusters (within 500 meters). When 
setting $min\_sample=20, radius=20\ meters$,  locations such as
highway transitions may fall within the same cluster.
The other combinations fail to cover a significant 
proportion of group event venues
in any of the group location clusters.
We further confirm that by choosing $min\_sample = 40, radius = 20\ meters$,  
the DBSCAN algorithm is able to find sensible clusters 
in groups' location traces based on a visual 
inspection of these clusters plotted on a map. 
These location clusters appear to correspond to locations frequented visited 
by our participants, such as work, school, home, etc.
Based on this analysis, 
we set the $min\_sample$ value as 40 and the $radius$ value as 20 meters
in our group location cluster detection task.

It is also worth noting that in roughly 10\% of the group events, 
the groups' final meeting venues did not fall into any of the group clusters detected. 
One reasonable explanation is that some 
group members may have been to this location some time ago but not recently. 
For example, users may have been to Costco two weeks ago, 
which is not covered by more recent location traces. 
It is also possible that groups of users decided to 
explore a new venue that they had never tried before. 
We plan to investigate this problem in more detail in the future when we collect
more data.

\section{Social-based Group Location Cluster Prediction Model}
\label{sec:prediction}

Given the group location clusters determined in the previous section, our goal
in this section is to design a statistical model to predict which location
cluster the group is going to meet at.  
Motivated by prior
work in group recommendation~\cite{gartrell2010enhancing,chaney2014large}, 
we divide this problem into two steps.
First, we analyze strong predictors that will impact group member's meeting decisions
and construct an innovative \modelname{Edge-RWR+CART} model 
to incorporate these identified features for predicting individual location preference.
After that, we aggregate group members' location preferences to predict group decisions.
Earlier study also suggests that the group decision strategy
varies depending on the strength of social connections
among group members~\cite{gartrell2010enhancing}. 
Inspired by this observation, in the context of group event venue recommendation, 
we propose a novel \modelname{social-based} group location 
prediction model, which adaptively applies different group decision strategies
to groups with different social relationship strength to aggregate group members'
individual location preferences.
	
\subsection{Overall Architecture of the Model}
\begin{figure}
\centering
  \includegraphics[width=1\linewidth]{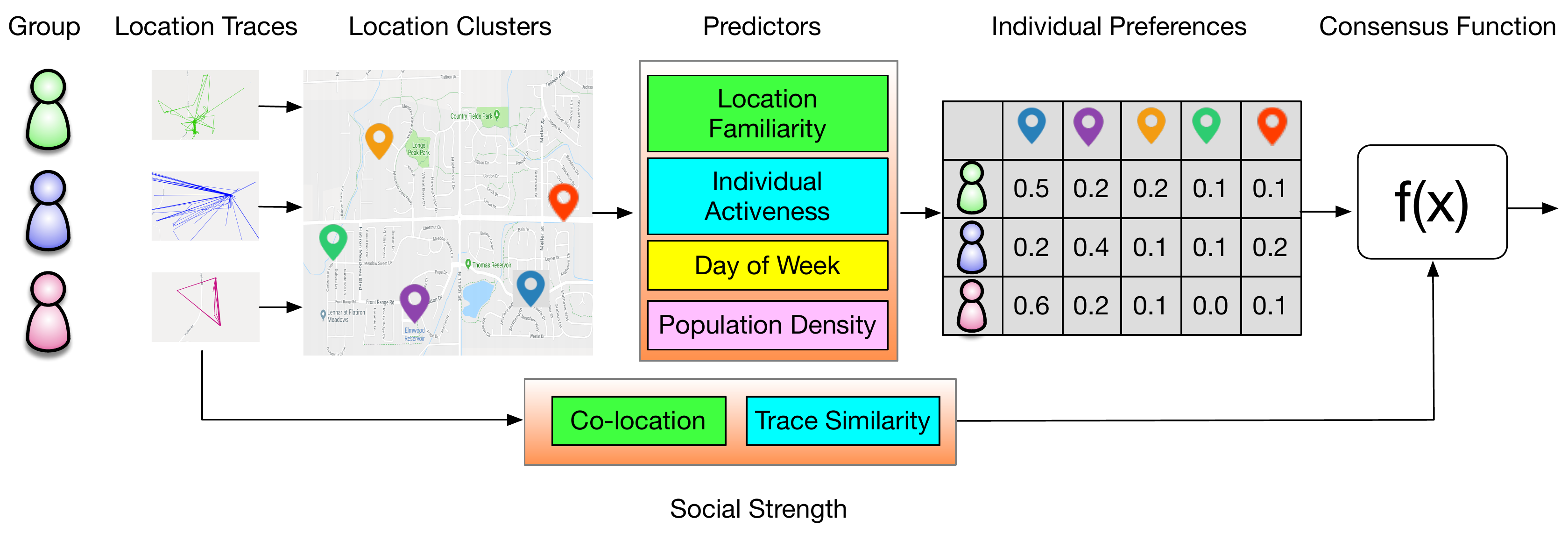}  
\caption{The overall architecture of our group location prediction model. 
With the detected group location clusters, 
we model each group member's individual location cluster preference based on his/her location traces.
Meanwhile, the location traces are also used for estimating the group's social relationship strength.
After that, a newly designed consensus function that we term a \modelname{social-based} group decision strategy 
is  used to aggregate individual members' location
cluster preferences and predict the group's final decision.}
\label{fig:predictionarc}
\end{figure}
The overall architecture of our prediction
model is illustrated in Figure~\ref{fig:predictionarc}.  As shown in the figure, 
individual group members' location traces are first merged to generate the group location
clusters shown on the map.  We then determine each individual's
likelihood of visiting each location cluster, as shown in the table on the
right side of the figure. This is accomplished by performing an analysis of
the key factors influencing each individual's likelihood of visiting a location
cluster, including location familiarity, individual activeness, day of the week,
and population density.  After determining the individual likelihoods of visiting a location cluster, we then
measure the social strength of the group in order to combine the
individual location preferences into a prediction of the group's
likelihood of visiting each location cluster. This is depicted as the group
consensus function $f(x)$ in the figure, along with the two social strength
factors that we will use, co-location and trace similarity.

\subsection{Social Relationship Strength}
\label{sec:social}
The social relationships among group members play a significant role in group
decision making.
Consider the following scenario: Tony and Paul are in the same group voting 
on a location for dinner on Friday night, and Paul does not
like spicy food.
If Tony knows Paul very well, he may not suggest
or vote for Thai restaurants that are famous for spicy cuisine, even though he himself 
loves spicy food.
A strong social relationship with Paul helps Tony to propose better meeting options. 
Here we estimate a group's social relationship strength using two factors:
\begin{itemize}
  \item \textit{Average number of daily meetings detected between any two group
  members using their location traces}: 
  The more often two users meet in real life, the more likely that they know
  each other well.
  We use the following criteria to define a meeting in our analysis:
  Two members must be approximately at the same location (within 20 meters)
  for at least five minutes.
  Please note that we use GPS location points, instead of location clusters, as a
  more accurate estimate of group members' meetings can be obtained using this
  data.  For example, two friends can coincidentally appear at the same farmers
  market for more than five minutes but not see each other, and thus this should
  not be counted as a meeting.
  \item \textit{Average similarity of group location cluster familiarity 
    between every pair of group members}: 
  Two users with more similar location cluster familiarity tend to know each
  other better.  Here we model the group location cluster
  familiarity of user $u$ by vector $\mathbf{loc}(u)$, where the $i$th element
  of $\mathbf{loc}(u)$ represents the number of location trace points in
  the $i$th location cluster detected in our dataset.  Furthermore,
  $\mathbf{loc}(u)$ is normalized per user.
\end{itemize}
Formally, we first measure the social relationship between two group members $u_i$ and $u_j$ as
\begin{equation}
\label{eq:social}
\text{social}(u_i, u_j)=w_1M(u_i, u_j)+(1-w_1)\cdot Cosine(\mathbf{loc}(u_i),\mathbf{loc}(u_j)),
\end{equation}
where $M(u_i,u_j)$ is the normalized average number of meetings per day detected between group
members $u_i$ and $u_j$, and $\mathbf{loc}(u_i)$ is the vector representing
${u_i}'s$ location familiarity for each group location cluster.
$w_1$ balances the relative importance of co-occurrence and location trace
similarity. Normally user location co-occurrence is a much stronger signal
than location trace similarity in indicating social relationship strength. Hence
$w_1$ is set to 0.8 in our experiments based on the observations from our
dataset.
The similarity of two group members' location cluster familiarity vectors is
measured by Cosine similarity:
\begin{equation}
Cosine(\mathbf{loc}(u_i), \mathbf{loc}(u_j))=\frac{\mathbf{loc}(u_i)\cdot \mathbf{loc}(u_j)}
{||\mathbf{loc}(u_i)||\times||\mathbf{loc}(u_j)||}.
\end{equation}

Given the social relationship strength computed between pairs of users, a group's
social relationship strength is defined as the
average of all the pairwise social relationships among the group members:
\begin{equation}
\label{eq:groupsocial}
\text{Social}(G) = \frac{2\sum_{u_i,u_j\in G} \text{social}(u_i, u_j)}{|G|(|G|-1)},
\end{equation}
where $|G|$ represents the group size.
\begin{figure}
  \centering
  \includegraphics[width=0.7\linewidth]{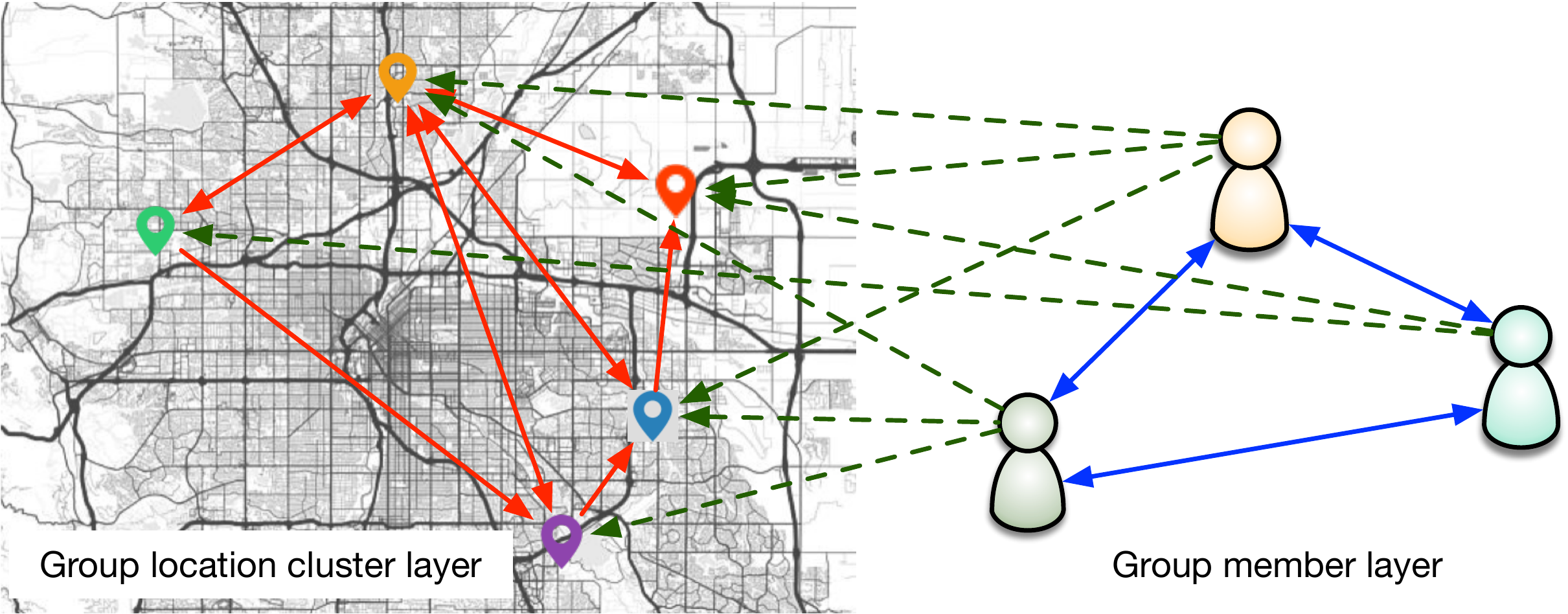}  
  \caption{An illustration of a group location cluster-group member graph.
    There are two layers in the graph: the group location cluster layer and the group member layer.
    The nodes in the location layer represent group location clusters, 
    while nodes in the group layer represent group members. 
    There are three types of edges in this graph: 
    the user-cluster edge (green), cluster-cluster edge (red)
    and user-user edge (blue).}
  \label{fig:randomwalk}
\end{figure}

\subsection{Location Familiarity Modeling for Individuals}
\label{sec:familiar}
Next, we focus on what factors may influence an individual to visit a particular 
location cluster.  Figure~\ref{fig:predictionarc} listed four such factors that we
considered, and in this section we focus first on location familiarity as a predictor,
followed by the three other factors considered in the next subsection.
The location familiarity of individuals within a group may vary from user to
user.
To model users' location familiarity more
accurately, we leverage transitions between group location clusters and the
social relationship between group members.
Intuitively, if a user's close friend is very familiar with one location
cluster, it is highly likely that this user has been there or at least heard
about it before, even though a visit to this location was not detected in  
location trace data.
Taking this into consideration, we construct a two-layer graph that effectively  
combines group members' social relationships and location trace information.
The structure of the graph is shown in Figure~\ref{fig:randomwalk}.
There are two layers in the graph:
the group location cluster layer and the group member layer.
The nodes in the group location cluster layer represent group location clusters,
and the nodes in the group member layer represent group members. There are
three types of edges in this graph: the user-cluster edge (green), the
cluster-cluster edge (red), and the user-user edge (blue).

\smallskip
\textbf{Edge-specified Random Walk with Restart (Edge-RWR).}
To estimate a group member's familiarity to a group location cluster,
we propose a novel Edge-specified Random Walk with Restart (Edge-RWR) algorithm
to calculate the relevance score between a group member and a group location
cluster.  Unlike the traditional RWR method that treats every edge equally, 
we define the transition probability of each edge follows.

The \textit{user-cluster} edge describes the probability that a user will visit
a location cluster.
Instead of simply using the number of visits, we apply the widely used idea of
TF-IDF \cite{ramos2003using}, where higher values indicate that the user is more
interested in this location cluster and this interest is significant among other
group members.
The same approach is utilized when estimating cluster-cluster and 
user-user edges' probabilities.
Specifically, given $m$ users and $n$ location clusters of a group, 
the user-cluster subgraph $\mathbf{G_{UC}}$ is represented by a 
$m \times n$ adjacency matrix $\mathbf{M_{UC}}$, where 
 $\mathbf{M_{UC}} = \{p(c_j|u_i)\}, 0\leq i<m, 0\leq i<n$, 
and $p(c_j|u_i)$ is estimated by:
\begin{equation}
p(c_j|u_i) = \alpha_1 \frac{Avg(u_i, c_j)}{Avg(c_j)} + (1-\alpha_1)\frac{1}{m},
\end{equation}
where $Avg(u_i, c_j)$ refers to the average number of visits per day 
made by user $u_i$ to cluster $c_j$. $Avg(c_j)$ refers to the average number
of total visits to cluster $c_j$ per day made by all the users in the group.
And $1-\alpha_1$ is the probability that the walker will restart from a 
randomly selected user node.

The \textit{cluster-cluster} edges measure the transition probabilities  
of users going from one location cluster to another. 
Given $n$ detected location clusters, the cluster-cluster subgraph 
$\mathbf{G_{CC}}$ is represented by a
$n \times n$ adjacency matrix $\mathbf{M_{CC}}$, where  
 $\mathbf{M_{CC}} = \{p(c_j|c_i)\}, 0\leq i<n, 0\leq j<n$, 
and $p(c_j|c_i)$ is estimated by:
\begin{equation}
p(c_j|c_i) = \alpha_2 \frac{Avg(c_i, c_j)}{Avg(c_j)} + (1-\alpha_2)\frac{1}{n},
\end{equation}
where $Avg(u_i, c_j)$ refers to the average number of transitions per day 
from cluster $c_i$ to cluster $c_j$. $Avg(c_j)$ refers to the average number
of total transitions to cluster $c_j$ per day from all of the group location
clusters.
And $1-\alpha_2$ is the probability that the walker will restart from a 
randomly selected location cluster node.

The \textit{user-user} edges represent   
how well two users know each other. 
We use $social(u_i, u_j)$, introduced in Equation~\ref{eq:social}, 
to model the social relationship strength between two users, $u_i$ and $u_j$.
Given $m$ group members, the user-user subgraph $\mathbf{G_{UU}}$ is represented by a
$m \times m$ adjacency matrix $\mathbf{M_{UU}}$, where 
 $\mathbf{M_{UU}} = \{p(u_j|u_i)\}, 0\leq i<m, 0\leq j<m$, 
and $p(u_j|u_i)$ is estimated by:
\begin{equation}
p(u_j|u_i) = \alpha_3 social(u_i, u_j) + (1-\alpha_3)\frac{1}{m},
\end{equation}
where $1-\alpha_3$ is the probability that the walker will restart from a
randomly selected user node.

\begin{algorithm}
\caption{Edge-specified Random Walk with Restart (Edge-RWR)}
\label{ag:edge}
\begin{algorithmic}[1]
\INPUT {Network $G = G_{UC} + G_{CC} + G_{UU}$;
		starting node $u_i$; restart probability $\alpha_1$, $\alpha_2$, $\alpha_3$;}
\OUTPUT Stationary group location cluster vector $w_C$ for a random walk starting at $u_i$;\\
Let all the nodes be initialized to 0, except  for a 1 for the $u_i$ node;\\
Let $w_U$ denote the weights of the column vector of users, and
$w_C$ denote the weights of the column vector of location clusters;\\
While ($\mathbf{U}$ and $\mathbf{C}$ have not converged):\\
\quad\quad${w_C}^{k+1} = \alpha_1\mathbf{M_{UC}}\cdot {w_U}^k + \alpha_2\mathbf{M_{CC}}\cdot{w_C}^k + (1-\alpha_2)e_C$;\\
\quad\quad${w_U}^{k+1} = (1-\alpha_1)e_U + \alpha_3\mathbf{M_{UU}}\cdot {w_U}^k + (1-\alpha_3)e_U $;\\
Output group location cluster vector $w_C$.
\end{algorithmic}
\end{algorithm}

\noindent\textbf{Time Complexity.} Algorithm~\ref{ag:edge} presents the
algorithm for finding the stationary vector of the random walk from a single
starting user node.
The time complexity of \modelname{Edge-RWR} is $O(k\cdot|V|^2)$, where $k$ is the number of
iterations required for convergence, and $|V|$ is the total number of
nodes.
According to \citet{tong2006fast}, the ratio of the first two eigenvalues of the
transition matrix specifies the rate of convergence to the stationary point.

After constructing the group location cluster-group member graph, 
\modelname{Edge-RWR} is then performed to compute the 
location familiarity between a user and a location cluster. 
The random walker will start from one node, and travel following
the probability assigned to each edge to select the next transition. 
In addition, in order to avoid stoppage of the random walk, we
assume that the random walker will in some cases jump back and restart at the same node. 
At each transition, the walker would either restart at one node (with
probability $1-\beta$), or move to a neighboring node (with probability $\beta$). 
In the two-layer graph that we constructed, the walker will
start from a group member node.
For each transition, the 
walker will either jump to an adjacent node, or jump back to the same starting
user node.
We set the parameter $\alpha_1=\alpha_2=\alpha_3 = 0.85$, 
which has also been widely used in prior literature. 
The intuition is that if a group location cluster node 
can be easily reached from a group member node, the user has a higher familiarity with this location cluster. 
Formally, the location familiarity 
of an individual group member $u$ for location cluster $c$ is modeled as
\begin{equation}
\label{eq:familiar}
\text{loc\_familiarity}(u, c) = \text{PageRank}_u(c),
\end{equation}
where $PageRank_u(c)$ is the PageRank score of group location cluster $c$ for
user $u$ after running the Edge-RWR algorithm starting from user $u$.

\subsection{Contextual Predictors}
\subsubsection{User Mobility}
User mobility can significantly impact daily activities. 
In a preference survey, 
Elgar \etal find that users with different levels of mobility differ
substantially in how they value time~\cite{elgar2004car}. 
Previous insights also suggest that users with high mobility (longer daily travel distance) tend to be more 
active in attending group events, 
and an individual's mobility has a positive correlation with her time
and location availability~\cite{zhang2018understanding}. 
There are two reasonable explanations for this phenomenon:
\begin{itemize}
\item As researchers have found, people with different modes of transportation,
such as walking, taking a bus, riding a bike, and driving, exhibit significant
difference in how they value time~\cite{elgar2004car,zheng2008learning}.
Users with high mobility may travel by car or public transportation.
Highly mobile users are more likely to attend group events that are far away
from their current locations.

\item Zhang~\etal also found that users with high mobility are more likely 
to be active event attendees, as they are used to frequent meetings  
with friends after school or work, 
which generates longer travel distances~\cite{zhang2018understanding}.
\end{itemize}
As such, we model the influence of user mobility as follows: 
a user's group event attendance is proportional to 
the user's mobility. 
Here we define user mobility as the total travel distance 
in the 48-hour period preceding the creation of the event.
More specifically, we estimate user mobility by summing up the geographical 
distance between every two consecutive location points detected 
from the user's location trace data in this time period.

\subsubsection{Day of the Week}
Using the detected group location clusters, 
we can identify a user's daily movement pattern among these location clusters. 
We find in our data analysis that users' movement patterns can be quite different
on weekdays compared with weekends.
The average inter-cluster movement distance is 5.3 km on weekends and 4.8 km on
weekdays.
This pattern seems reasonable, 
since intuitively people are more willing to travel longer distances for activities on weekends. 
And the average movement number is 5.2 on weekdays and 5.4 on weekends. 

\begin{figure}
\centering
\begin{subfigure}{0.5\textwidth}
  \centering
  \includegraphics[width=1\linewidth]{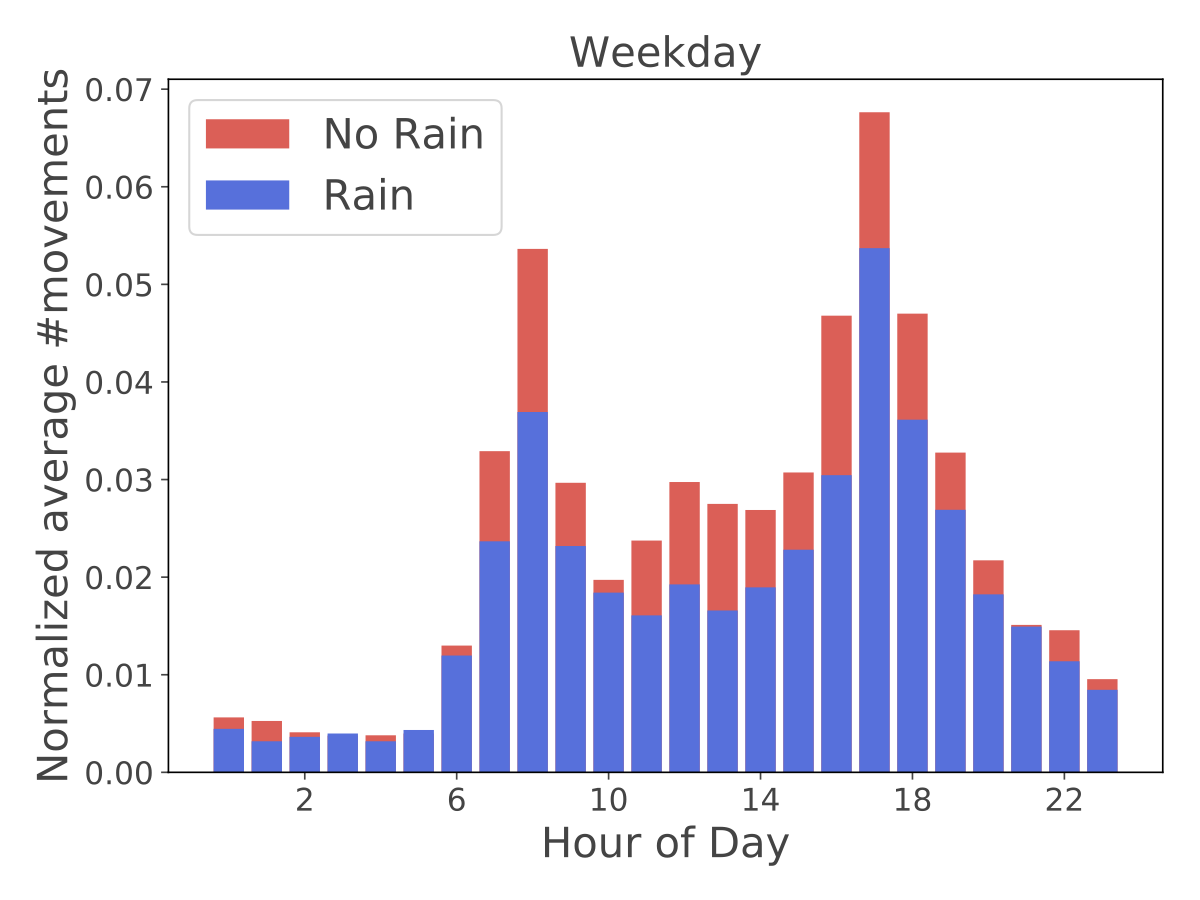}
  \caption{Number of movements by hour on weekdays.}
  \label{fig:weatherweekday}
\end{subfigure}%
\begin{subfigure}{0.5\textwidth}
  \centering
  \includegraphics[width=1\linewidth]{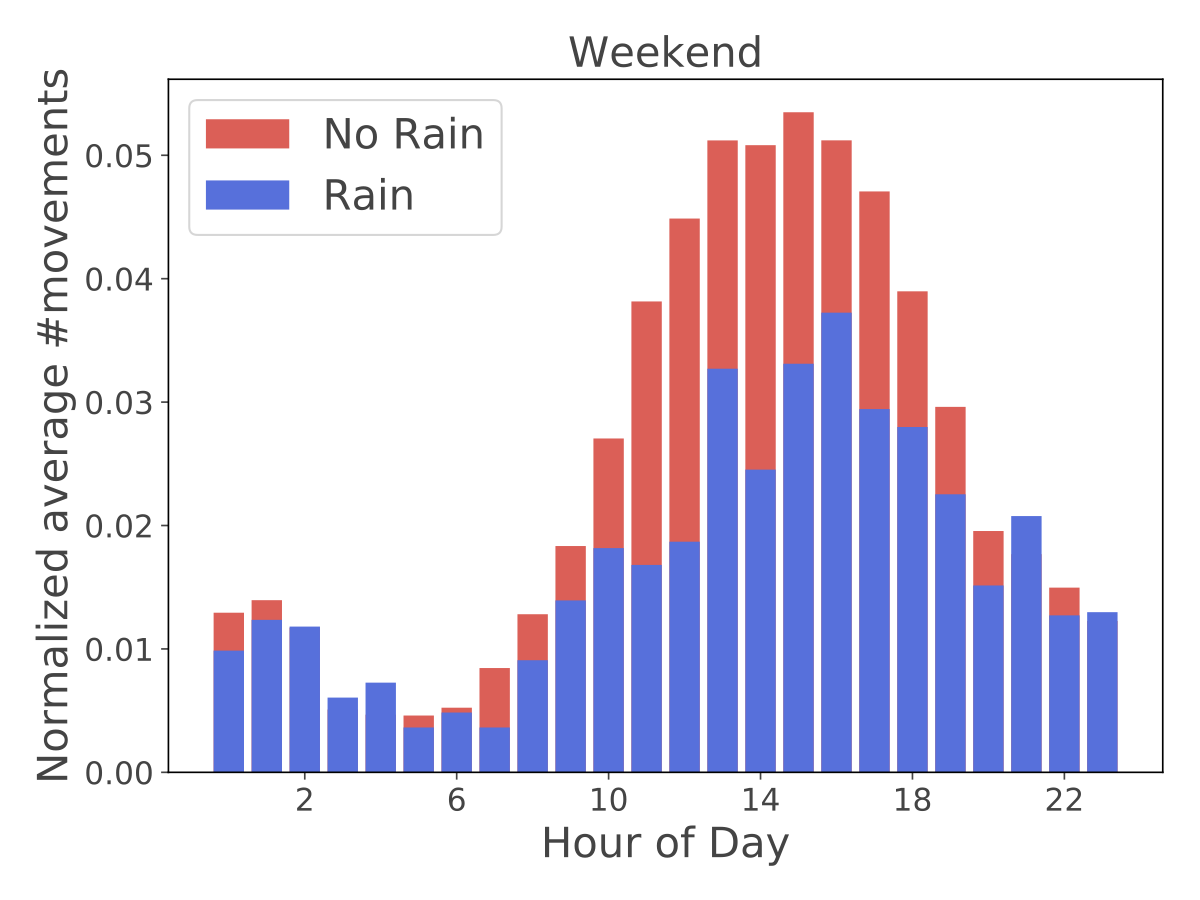}  
  \caption{Number of movements by hour on weekends.}
  \label{fig:weatherweekend}
\end{subfigure}
\caption{Figure~\ref{fig:weatherweekday} and~\ref{fig:weatherweekend} 
show the normalized average number movements on weekdays and weekends. 
Bad weather causes a sharp decrease of daily movements on both weekdays and
weekends, and the impact is larger on weekends compared with weekdays.}
\label{fig:weather}
\end{figure}

We also find that weather 
has a significantly different impact on weekdays compared with weekends.  
We control geographical confounding factors relating to weather 
by only using a partial dataset collected 
at our university. 
When obtaining the weather information during the time of a
given movement made by a user, we round the time to the nearest hour.  
For example, if the starting time of the
movement is at 18:15, we collect the weather data at 18:00 that day using the Weathersource API~\cite{weathersource}.
For the hours with positive precipitation values (including snow), we consider it to be rain. 
The normalized average movements by hours under different weather conditions 
on weekdays and weekends are shown in Figure~\ref{fig:weather}.
These results demonstrate that bad weather causes a sharp decrease in daily
movements on both weekdays and weekends.
We also find that the movement difference between rain and no-rain hours is
larger on weekends, indicating that weather has a higher impact on weekend
activities compared with weekdays.

In summary, users' movement patterns are considerably different on weekdays and
weekends, which also impact users' group meeting preferences. 
Therefore, it would be beneficial to consider the day of  the week information in our 
prediction model. 
We also considered including weather information as a contextual feature 
in our prediction model. 
However, since we are trying to recommend venues when a group event is
created,  %
only weather forecast information can be used, 
which can be inaccurate but also difficult to collect 
as our users are spread throughout the US. 
To ensure that our GEVR system is applicable in real-life cases, 
we decided not include weather forecast data in our recommendation system in this
paper.
However, incorporating the impact of weather forecasts on group event decisions
could be an interesting topic, which we save for future work. 

\subsubsection{Population Density}
\label{sec:density}
As shown in Figure~\ref{fig:map}, our users are spread all over the US, 
across 40 states and 117 cities.
Population density can play a significant role in modeling group members'
location preferences.
Previous research has found that geography and social relationships are inextricably intertwined.
People living in rural and urban places exhibit significant differences in using social 
technologies, such as Facebook and Myspace~\cite{gilbert2008network,backstrom2010find}.
These findings inspire us to study the effect of population density on group event scheduling.
Therefore, we also incorporate the population density information into our prediction model.

\subsection{Individual Location Preference Prediction}
Taking all of the features we have discussed thus far into consideration, 
we integrate all of the contextual features into a Classification and Regression
Tree (CART) model \cite{loh2011classification}. 
We predict a user's location preference with four types of features: 
\textit{location familiarity}, \textit{user mobility}, 
\textit{day of the week}, and \textit{population density}. 
Location familiarity is calculated using Equation~\ref{eq:familiar} 
introduced in Section~\ref{sec:familiar}. 
User mobility models a user's average travel distance on a daily basis. 
For users with high mobility, traveling longer distances can be easier
compared with low-mobility users, due to the fact that high-mobility users
generally travel by car or public transportation, and they are accustomed to
longer-distance travel.
Our analysis further illustrates that users' movement patterns vary
significantly on weekdays versus weekends, and in high population density areas
versus low population density areas; thus, we also add these two features into
our regression model.
We also evaluated other popular regression methods, such as Linear Regression, Support Vector
Regression, and Gradient Boosting Regression. 
CART outperforms all of these regression models for our application. We
present detailed results in Section~\ref{sec:evaluation}.

One major advantage of our individual location preference prediction model is its flexibility
in leveraging new features. Higher-level mobility features, which are related to
users' group event attendance activity, can be easily added to our current
model.
For example, \citet{stenneth2011transportation} and \citet{hemminki2013accelerometer} have 
found that a user's mode of transportation, 
including car, bus, train, walking, and bike, can be accurately inferred
using GPS location traces. 
As future work, we plan to investigate how users' modes of transportation can impact group
event attendance.

\subsection{Social-based Group Location Prediction Model}
\label{sec:group}

Thus far, we have constructed a prediction model 
that integrates all the features we have identified in an intelligent way
to model an individual's location preferences. 
Now, we propose a new \modelname{social-based} group location prediction model
to aggregate the predicted location preferences for individuals
and predict the final location cluster where the group will meet. 

Given every group member's location preferences, 
how does the group arrive at a final decision of where to meet? 
In previous group recommendation work, 
three main strategies for aggregating individual preferences 
have been widely explored~\cite{jameson2007recommendation,gartrell2010enhancing,chaney2014large}: 
\modelname{least misery,} \modelname{average satisfaction,} and \modelname{maximum satisfaction.}
\begin{itemize}
\item \textit{Least misery} minimizes the dissatisfaction of the least satisfied
member:
$\min_{u\in G}\mathbf{p}(u, c)$.
\item \textit{Average satisfaction} is a straightforward strategy 
that assumes every group member carries the same weight and 
computes the average satisfaction: $\frac{1}{|G|}\sum_{u\in G}\mathbf{p}(u, c)$.
\item \textit{Maximum satisfaction} maximizes the enjoyment of the most
satisfied group member:
$\max_{u\in G}\mathbf{p}(u, c)$.
\end{itemize}
Earlier studies have also argued that groups are diverse and 
none of these group decision strategies are dominant across all 
groups~\cite{masthoff2004group,gartrell2010enhancing}. 
Gartrell \etal found that the social relationship strength of a group 
plays a significant role in the group decision-making process. 
When the relationship strength is high, the final decision tends to reflect
the maximum satisfaction strategy.
When the relationship strength is weak, the final decision tends to reflect average satisfaction 
or least misery~\cite{gartrell2010enhancing}. 
The intuition here is clear: If the group members don't know each other, 
the main guidance for making the group decision is to avoid dissatisfying the
other group members.
If the group has high social relationship strength, some people who are knowledgeable 
about the topic may lead the discussion and others tend to follow their
suggestions.
Taking these factors into consideration, 
we propose the following \modelname{social-based} group consensus function:
\begin{equation}
\mathbf{P}(G, c) = \left\{ \begin{array}{rcll}
min_{u\in G}\mathbf{p}(u, c) & \mbox{if}
& \text{Social}(G)<\beta & \mbox{(least misery)} \\ 
\frac{1}{|G|}\sum_{u\in G}\mathbf{p}(u, c) & \mbox{if} & \beta\leq\text{Social}(G)\leq\alpha & \mbox{(average satisfaction)} \\
max_{u\in G}\mathbf{p}(u, c) & \mbox{if} & \alpha<\text{Social(G)} &
\mbox{(maximum satisfaction)}, \end{array}\right.
\label{eq:pred}
\end{equation}
where $\text{Social(G)}$ is defined in Equation~\ref{eq:groupsocial}, and
parameters $\alpha$ and $\beta$ are thresholds for the group's social
relationship strength.
We choose $\alpha = 0.6$ and $\beta = 0.2$ in our evaluation based on a grid search targeting the best performance
for predicting groups' final meeting location clusters using our dataset.
We will present the details of our experiments in
Section~\ref{sec:evaluation}.

\section{Venue Recommendation Framework}
\label{sec:recommender}
\begin{figure}
\centering
  \includegraphics[width=0.85\linewidth]{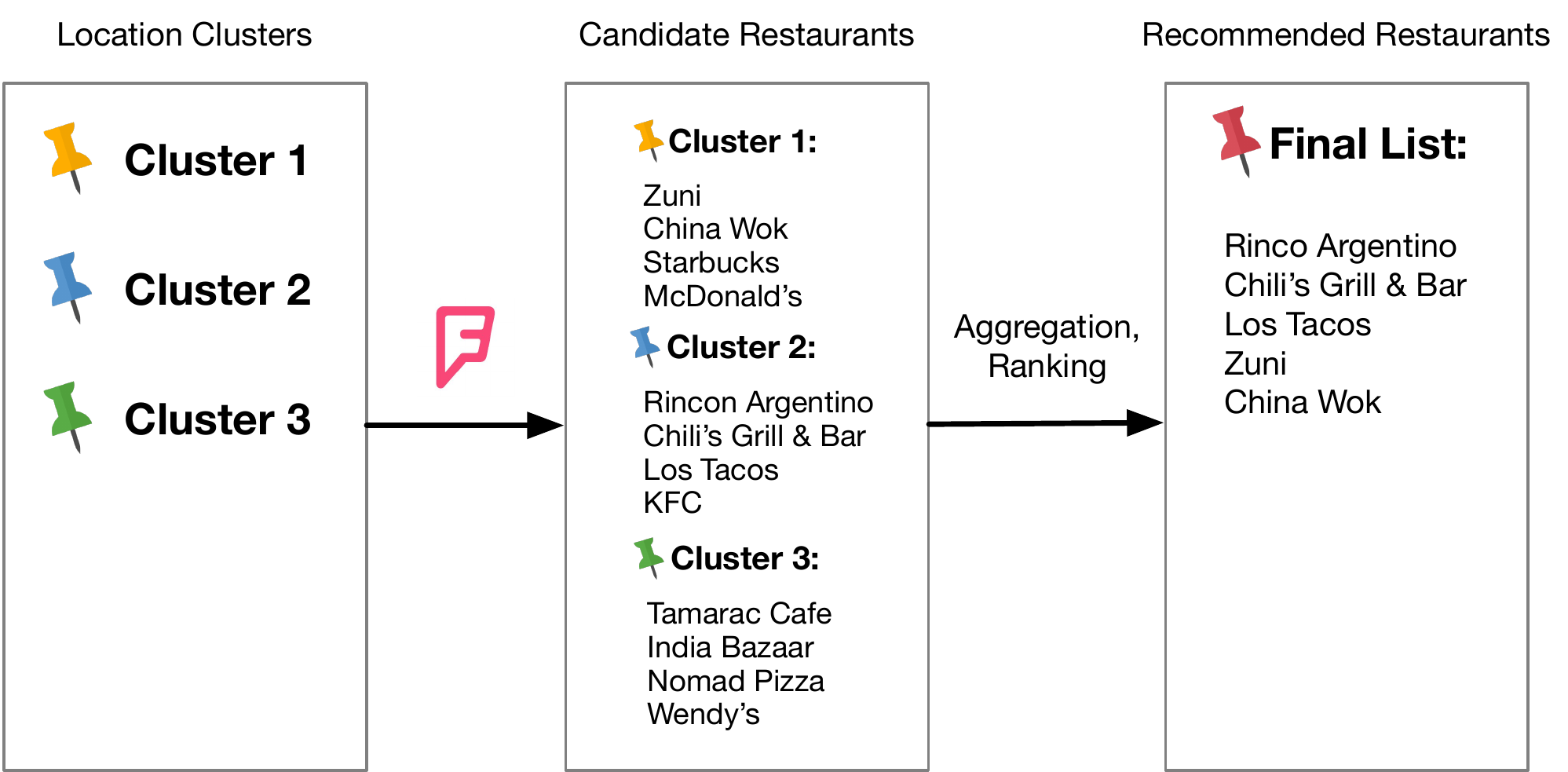}  
\caption{The main flow of the group event venue recommendation framework: 
For each detected group location cluster, 
we call the Foursquare Venue Recommendation API to get a list of nearby restaurants. 
We aggregate the restaurants of each cluster to generate a candidate restaurant pool. 
Prediction results from the \modelname{social-based} group location cluster prediction model are then used
to rank the restaurant pool and generate the final Top-K restaurant list.}
\label{fig:reco}
\end{figure}
In this section, we present our proposed group event venue recommendation framework. 
The intuition behind our framework is to  re-rank group's nearby venues, 
returned by Foursquare's API, based on the group location cluster prediction results 
calculated by Equation~\ref{eq:pred}
to provide event venue recommendation for groups. 
The main flow of the framework is shown in Figure~\ref{fig:reco}.
Aiming to recommend event venues for groups of mobile users in real-life,
our framework is composed of three parts:
\begin{enumerate}
\item The group location clusters generated 
by grouping location traces' points described in Section~\ref{sec:cluster}; 
\item The \modelname{social-based} group location prediction model 
introduced in Section~\ref{sec:prediction}; 
\item The venue recommendation system 
provided by Foursquare's API~\cite{foursquare}.
\end{enumerate}

Given a combination of location coordinates, a radius, 
and a limit (maximum number of results to return),
the Foursquare API will return a list of recommended venues near the specified location within the given radius. 
In our setting, given a group $G$ and its group location clusters $\{c_1, c_2, \cdots, c_n\}$, 
we call the Foursquare API to return a list of $m$ recommended venues $\{v_{i1}, v_{i2}, \cdots, v_{im}\}$ 
for each cluster $c_i$. 
Here $\{v_{i1}, v_{i2}, \cdots, v_{im}\}$ follows the order returned by the Foursquare API.
All recommended venues are combined to generate a candidate venue pool. 
Two factors are leveraged to re-rank the venue pool: 
the venue rank in the returned list provided by the Foursquare API 
and probabilities of corresponding location clusters being selected 
as the final one calculated by our \modelname{social-based} group location prediction model. 
More specifically, the rank score of a venue ${v_{ij}}$ 
in the final recommendation list is defined as
\begin{equation}
\label{eq:rec}
\text{rankscore}(v_{ij}) = \frac{m-(j-1)}{m} \cdot \mathbf{P}(G, c_i),
\end{equation}
where $\frac{m-(j-1)}{m}$ models the rank weight
returned by the Foursquare API and $\mathbf{P}(G, c_i)$ 
is the probability of location cluster $c_i$ being selected 
as the final location cluster computed by Equation~\ref{eq:pred}. 
The final venue recommendation list created by our framework 
is ranked by $\text{rankscore}(v_{ij})$ 
among all the venues in the candidate venue pool. 

\section{Evaluation}
\label{sec:evaluation}
In this section, we evaluate the effectiveness of GEVR's proposed 
individual location preference prediction model,
social-based group location cluster prediction model, 
and group event venue recommendation framework.

\subsection{Performance of Predicting Individual Location Preference}
We first evaluate the performance of predicting individual location preference. 
Specifically, we evaluate ``whether a group member will vote for 
a suggested venue or not,'' and the answer is a binary label 
with 1 representing \textit{yes} and 0 representing \textit{no}.
The evaluation metrics we use are AUC (the area under the ROC curve) and
F1 score. Since our model quantifies a group member's location cluster preference 
by outputting a probability value, 
AUC can find the best threshold for the binary classification. 
And F1 score is computed to evaluate 
whether the model provides comparatively robust performance. 
We use a stratified 5-fold cross validation with respect to the users. 
All the features are standardized. 

As introduced in Section~\ref{sec:prediction}, our prediction model utilizes four main types of features: 
location familiarity, user activeness, day of the week, and population density. 
By comparing the performance of the popular supervised models shown in 
Table~\ref{tab:individual}, we find that \modelname{Edge-RWR+CART} achieves the best results. 
The other supervised regression models include Linear Regression (LR), 
Support Vector Regression (SVR), and Gradient Boosting Regression (GR).
\modelname{Edge-RWR+CART} achieves the best overall performance in estimating users' location
cluster preference (AUC = 0.851, F1 = 0.831) and  
outperforms all the other supervised models by a clear margin (beating
the second best one \modelname{Edge-RWR+GR} by 8.9\% in AUC and 7.1\% in F1 Score).
More importantly, our newly proposed \modelname{Edge-RWR} beats the traditional \modelname{RWR}
consistently when combined with each supervised learning model  
that we tested with an average improvement of 13\%, indicating that 
our edge-specified network is superior in estimating the location cluster preference 
between a group member and a location cluster.

%

\subsection{Performance of Predicting Where Groups Meet}

\begin{table}[]
\centering
\caption{Performance for predicting individual location preference. Standard error values are shown in parentheses.
\modelname{Edge-RWR+CART} outperforms all the other supervised models by a clear margin. 
Moreover, our newly proposed \modelname{Edge-RWR} beats the traditional \modelname{RWR}
consistently when combined with each supervised learning model that we tested, 
indicating that our edge-specified network is superior 
in estimating the familiarity between a group member and a location cluster.}
\begin{tabular}{ccc}
\hline
\textbf{Model}                                 & \textbf{AUC}           & \textbf{F1 Score}            \\ \hline
RWR+LR            & 0.702 (0.020)     & 0.675 (0.021) \\
Edge-RWR+LR           & 0.705 (0.017)     & 0.669 (0.022) \\
RWR+SVR           & 0.745 (0.010)     & 0.722 (0.008) \\
Edge-RWR+SVR          & 0.763 (0.011)     & 0.757 (0.009) \\
RWR+GR            & 0.788 (0.015)     & 0.747 (0.013) \\
Edge-RWR+GR           & 0.795 (0.011)     & 0.766 (0.008) \\
RWR+CART          & 0.838 (0.011)     & 0.812 (0.009) \\
\textbf{Edge-RWR+CART}   & \textbf{0.851  (0.011)} & \textbf{0.831  (0.010)} \\ \hline
\end{tabular}
\label{tab:individual}
\end{table}

The next question is whether we can accurately predict where a group meets 
by using our newly designed \modelname{social-based} group decision strategy. 
We formulate the problem as a \textbf{multi-class prediction task}: Given all the group location
clusters, can we accurately predict which location cluster the group will decide to meet at.
Note that the design of the mobile application allows a host to override the voting results by selecting a final venue 
that is not necessarily the winning one (based on group members' votes). 
This may happen when group members discussed where to hang out and 
made an agreement to disregard the poll results. 
In our dataset, the host decided to override the majority voting results in roughly 20\% of the group events. 
Therefore, we could measure the effectiveness as either 
predicting the most popular location cluster according to the poll results, 
i.e., the ``winning location cluster,'' 
or predicting the final location cluster actually visited by the group including overrides, namely the ``final location cluster.''  To evaluate the performance of our model, 
we report the prediction accuracy of both the winning location cluster and the final location cluster. 
We compare our \modelname{social-based} group location prediction model
with the three commonly used group decision models introduced in Section~\ref{sec:group}, 
and three other state-of-the-art group recommendation models that have been shown to be effective 
in their specific group settings in prior literature:
\begin{itemize}
\item \textit{Logsitic regression-based} \cite{chaney2014large}: A logistic regression-based approach that 
determines the probability of a group view ($p_G$) from the individual probabilities:
\begin{equation}
\log\frac{p_G}{1-p_G} = \alpha_0 + \alpha_1p_1 + \alpha_2p_2 + \alpha_3p_3 + ... + \alpha_np_n.
\end{equation}
\item \textit{Rule-based} \cite{gartrell2010enhancing}: A rule-based group consensus framework that constructs
associative classification rules by mining the training dataset. 
\item \textit{Expertise-based} \cite{quintarelli2016recommending}: Estimating group decision by averaging 
the individual preference with weighted expertise. The expertise is defined as the number 
of times this user participated before. 
\end{itemize}

\begin{table}[]
\centering
\caption{Accuracy of predicting the winning location cluster and final location cluster. Individual location preferences are estimated using \modelname{Edge-RWR+CART} as it achieves the highest performance in modeling individual location preference. Standard error values are shown in parentheses. Our \modelname{social-based} group location prediction model 
achieves better prediction accuracy than all the other comparative models,
with an average improvement of 12\%.}
\label{tab:group}
\begin{tabular}{ccc}
\hline
\textbf{Model}                                 & \textbf{Winning location cluster}           & \textbf{Final location cluster}            \\ \hline
Least misery \cite{quintarelli2016recommending}			& 0.689 (0.019)		& 0.668 (0.018)	\\
Average satisfaction \cite{gartrell2010enhancing,chaney2014large,quintarelli2016recommending}	& 	0.771 (0.016)	& 0.753 (0.017)	\\
Most pleasure \cite{chaney2014large,quintarelli2016recommending}			& 0.760 (0.019)		& 0.739 (0.022)	\\
Logistic regression-based	\cite{chaney2014large}		& 0.767 (0.016)		& 0.741 (0.019)	\\
Rule-based \cite{gartrell2010enhancing}			& 0.686 (0.022)		& 0.657 (0.024)	\\
Expertise-based \cite{quintarelli2016recommending}			& 0.758 (0.018)		& 0.729 (0.016)	\\
\textbf{Social-based}                         & \textbf{0.826  (0.013)} & \textbf{0.801  (0.013)} \\ \hline
\end{tabular}
\end{table}
The results of predicting the winning location cluster 
and the final location cluster are shown in Table~\ref{tab:group}. 
As can be observed in the table, 
our newly proposed social-based group decision strategy achieves the best performance 
in both predicting winning location cluster (Accuracy=0.826) and final location cluster (Accuracy=0.801).
It is also worth noting that our \modelname{social-based} group location prediction model 
achieves better prediction accuracy than all other three widely-used single group decision strategies,
with an average improvement of 12\%. 
This echoes the insights illustrated in~\cite{gartrell2010enhancing,masthoff2004group}: 
groups are diverse and none of the single group decision strategies are dominant across all groups.
The accuracy achieved by the \modelname{logistic regression-based} approach also shows that a linear combination
of individual preferences without considering the social relationship strength among group members
does not perform well. 
Moreover, our \modelname{social-based} model performs much better than the 
\modelname{expertise-based} model. One possible explanation 
is that our dataset currently does not have enough events created by the same group. 
As a result, the expertise of group members cannot be accurately modeled in such cold-start scenarios. 
We would like to explore if the \modelname{expertise-based} model will perform better when we have 
more group events data. We save this topic for future study.

{\bf Performance for groups of different sizes.} 
To better understand the impact of group size on recommendation performance, we computed the prediction accuracy for different group sizes using our \modelname{social-based} group location cluster prediction model. Our model achieves an average accuracy of 0.818 for groups with three members (0.826 for the winning location cluster and 0.810 for the final location cluster),
0.807 for groups with four members (0.833 for the winning
location cluster and 0.781 for the final location cluster),
and 0.806 for groups with 5--6 members (0.823 for the winning
location cluster and 0.788 for the final location cluster).
These results show that our model performs reasonably well for
groups with sizes between 3--6. However, the trend or the change in prediction performance is not statistically significant. As described in Section~\ref{sec:data}, only 17\% of the groups in our dataset have more than four members. This would be an interesting problem to explore when we collect more data for larger groups.

\begin{figure}
    \begin{subfigure}[t]{0.48\textwidth}
        \includegraphics[width=\textwidth]{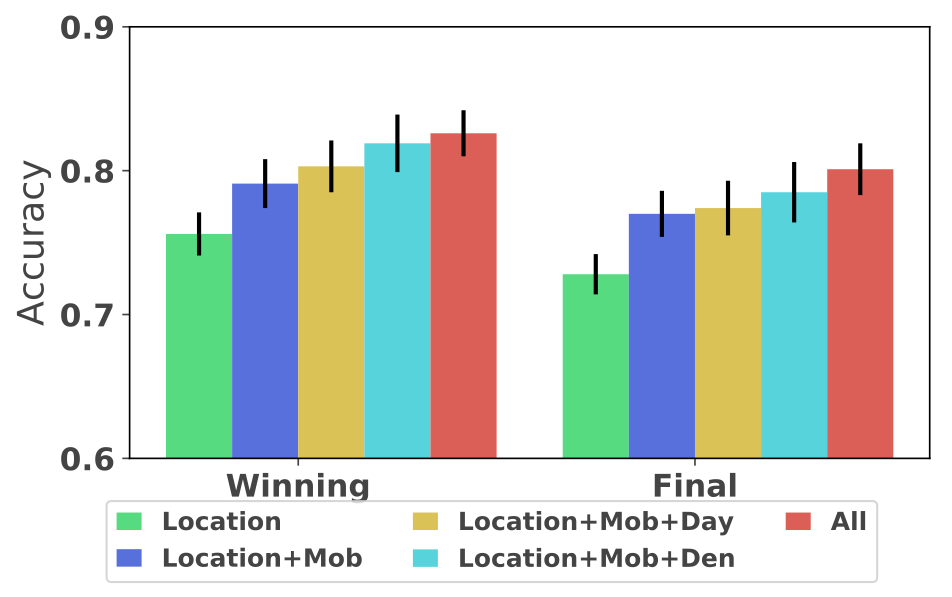}
        \caption{Feature analysis.}
        \label{fig:feature}
    \end{subfigure}
    \hfill
    \begin{subfigure}[t]{0.48\textwidth}
        \includegraphics[width=\textwidth]{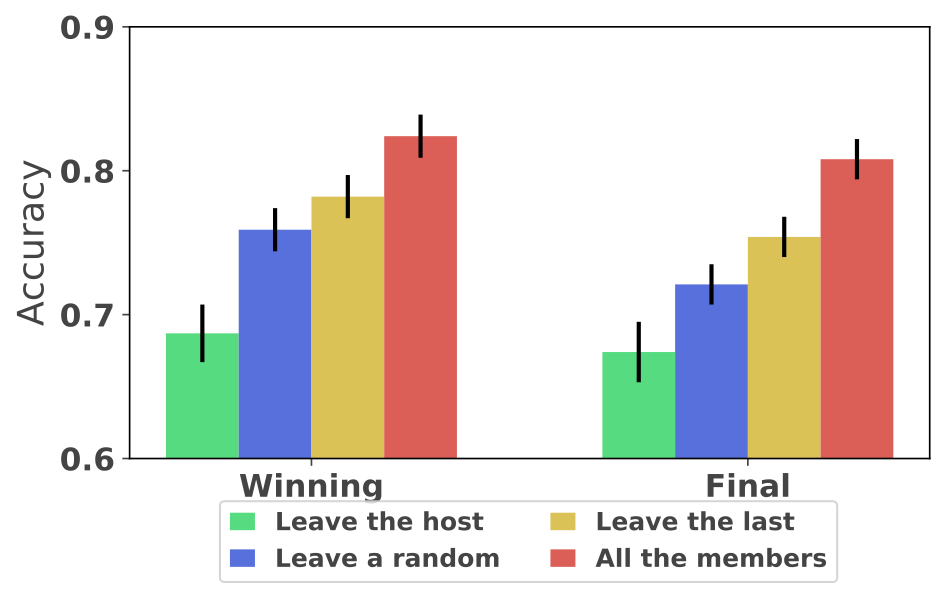}
        \caption{Cold-start analysis.}
        \label{fig:coldstart}
    \end{subfigure}
    \caption{Feature analysis (Figure~\ref{fig:feature}) and cold-start analysis  
    (Figure~\ref{fig:coldstart}). In Figure~\ref{fig:feature}, using the location
    familiarity and user mobility features already provide reliable performance. 
    Both day of the week and population density features also contribute
    to the overall improvement. In Figure~\ref{fig:coldstart}, the group
    host's location information has the most significant impact on the overall
    prediction performance when predicting both the ``winning location cluster''
    and ``final location cluster,'' while the last joined group member has the least
    impact.}
    \label{fig:analysis}
\end{figure}

{\bf Feature analysis.}
Our \modelname{social-based} prediction model utilizes four types of features: Location familiarity (Location), user mobility (Mob), day of the week (Day), and population density (Den). To understand the effectiveness of these features, we evaluate the prediction performance using different feature combinations. As shown in Figure~\ref{fig:feature}, using location familiarity alone achieves reasonable performance, and adding user mobility 
further increases the accuracy (from 0.756 to 0.791 for predicting the winning location cluster, and from 0.728 to 0.770 for predicting the final location cluster). 
\added{One possible explanation is that users with high mobility may travel by car or use public transportation.
Reaching far-away places are thus easier for them. 
It is also likely that users with high mobility are inherently active event attendees, as they are used to
frequent meetings with friends after school or work~\cite{zhang2018understanding}.}
Additionally, both day of the week and population density features contribute to the overall improvement.

{\bf Performance on Groups with Cold-Start Users.}
As we discussed in previous sections, it is common for groups to have newly joined members 
without (sufficient)  historical information. 
To evaluate our \modelname{social-based} group location prediction model in such cold-start scenarios, 
we set up two experiments:
\begin{itemize}
	\item \textit{Leave-one-out}: Assuming one group member is a new user of the app, we  
	do not have any historical information about him/her.
	\item \textit{Effect of data sparsity}: We further study how our model deals with the cold-start problem
	by using only a proportion of each user's data (20\%, 40\%, 60\%, 80\%, and 100\%). 
\end{itemize}
Figure~\ref{fig:coldstart} shows the performance of the ``leave-one-out'' experiment. 
We simulate three ``leave-one-out'' situations by leaving (a) the group host, (b) a random group member 
(except the host and the last joined group member), or (c) the last joined group member as the new user 
with no historical information available. 
As we can see in Figure~\ref{fig:coldstart}, the group host's location information has the 
largest impact on the overall prediction performance when predicting both the ``winning location cluster''
and the ``final location cluster,''  while the last joined group member has the least impact. 
This finding echoes an observation made in \cite{zhang2018understanding}: ``The group host will have more 
influence on the group decision-making process. 
The late coming voters tend to vote for options that 
align with existing voting results, thus, having smaller impacts on the final decisions.'' 
More importantly, our \modelname{social-based} group location prediction model achieves promising results with cold-start users in the group, with an average accuracy of 0.768 when assuming the last group member' data is unknown and
an average accuracy of 0.740 when leaving a random group member's data out.

\begin{figure}
    \begin{subfigure}[t]{0.48\textwidth}
        \includegraphics[width=\textwidth]{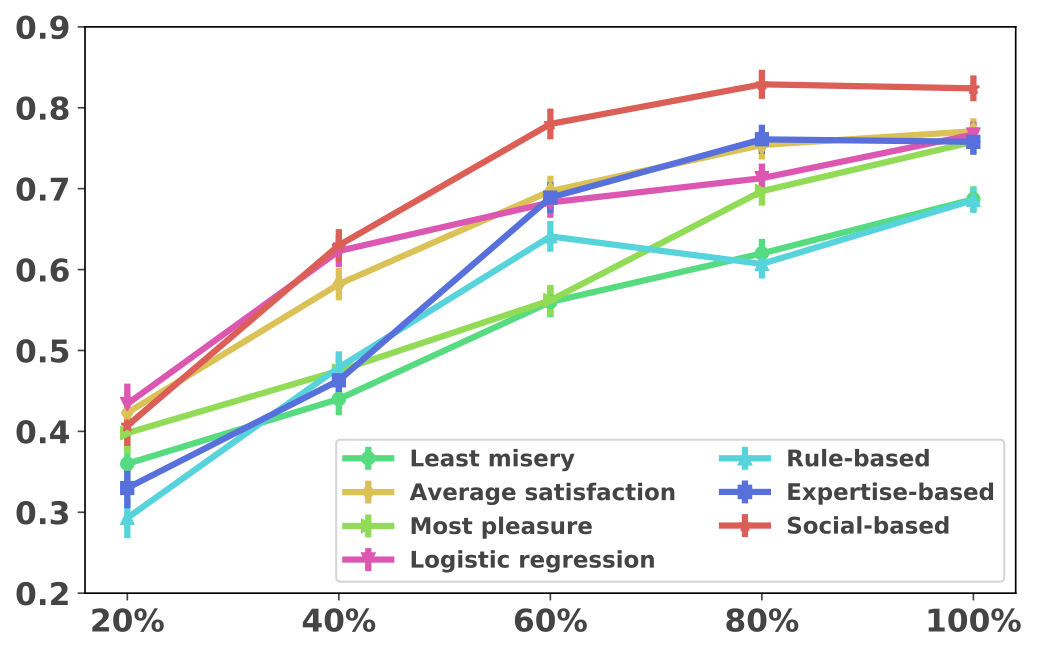}
        \caption{Accuracy of predicting the winning location cluster.}
        \label{fig:winning}
    \end{subfigure}
    \hfill
    \begin{subfigure}[t]{0.48\textwidth}
        \includegraphics[width=\textwidth]{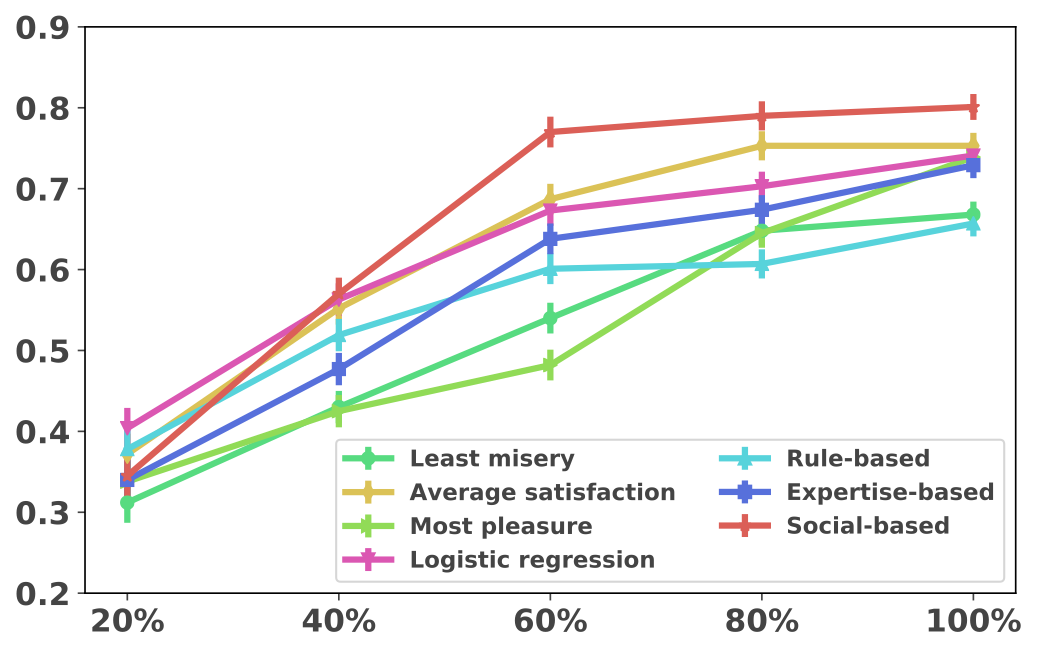}
        \caption{Accuracy of predicting the final location cluster.}
        \label{fig:final}
    \end{subfigure}
    \caption{Accuracy of predicting the winning location cluster (Figure~\ref{fig:winning}) and the final
    location cluster (Figure~\ref{fig:final}) using all the models in Table~\ref{tab:group} 
    with different proportions of users' location trace data. Our proposed \modelname{social-based} model consistently
	outperforms all the other models when the proportion of data is over 40\%.}
    \label{fig:spasity}
\end{figure}

We further study how our proposed model deals with the data sparsity problem by using different proportions 
of the users' location traces and compare it with all the other models shown in Table~\ref{tab:group}. 
The results are shown in Figure~\ref{fig:spasity}. 
We gradually increase the proportion of users' location traces used for testing the 
performance from the first 20\%, first 40\%, till 100\% (temporally ordered). 
As can be seen in the figure, our proposed \modelname{social-based} model consistently
outperforms all the other models when the proportion is above 40\%. 
It does not perform very well when only 20\% of the 
location traces are available for prediction. It is likely that the social relationship strength 
among group members cannot be precisely estimated with very limited data, 
which affects the overall prediction accuracy.

\subsection{Performance of Event Venue Recommendation for Groups of Mobile Users}
To evaluate whether our proposed~\systemname{} system is effective for providing event venue 
recommendations for groups of mobile users, we compare it with the following baselines: 
\begin{itemize}
\item \textit{Foursquare}: Get 50 recommended restaurants provided by the Foursquare Venue
Recommendation API by searching the city name where the group is located at and retain the 
restaurants' Foursquare rankings without any modification.
\item \textit{Most popular}: Same as the Foursquare model but re-ranks the restaurants by  
Foursquare number of check-ins (popularity).
\item \textit{Highest rating}: Same as the Foursquare model, but re-ranks the restaurants by  
Foursquare user review star rating. If there is a tie, the restaurant with more check-ins is ranked higher. 
\item \textit{Equal-weighted}: For each detected group location cluster, getting 50 
recommended restaurants by searching the nearby area within 0.5 km using the Foursquare Venue
Recommendation API and rank them using Equation~\ref{eq:rec}, assuming every location
cluster is weighted equally. If there is a tie, restaurants with more check-ins are ranked higher.
\item \textit{Least misery} \cite{quintarelli2016recommending}: 
Using the \modelname{least misery} group location prediction model to recommend venues.
\item \textit{Average satisfaction} \cite{gartrell2010enhancing,chaney2014large,quintarelli2016recommending}: 
Using the \modelname{average satisfaction} group location prediction model to recommend venues.
\item \textit{Most pleasure} \cite{chaney2014large,quintarelli2016recommending}: 
Using the \modelname{most pleasure} group location prediction model to recommend venues.
\item \textit{Logistic regression-based} \cite{chaney2014large}: 
Using the \modelname{logistic regression-based} group location prediction model to recommend venues.
\item \textit{Rule-based} \cite{gartrell2010enhancing}: 
Using the \modelname{rule-based} group location prediction model to recommend venues.
\item \textit{Expertise-based} \cite{quintarelli2016recommending}: 
Using the \modelname{expertise-based} group location prediction model to recommend venues.
\end{itemize}
Here, getting 50 recommended restaurants means setting the field \textit{limit} = 50 
as the number of results to return when using the Foursquare API. 
We choose 50 because it is the maximum number allowed. 
The returned list could be shorter than 50 if there are not enough restaurants registered nearby. 

The three basic baselines \modelname{most popular,} \modelname{highest rating,} 
and \modelname{Foursquare} are intended to demonstrate the recommendation performance 
without leveraging the detected group location clusters.
The \modelname{equal-weighted} baseline uses group location clusters information, 
but assumes no prediction results are provided, so all detected group location clusters are equally weighted.
The next six baselines combine individual location preference with different group decision strategies, aiming
to determine the feasibility of our social-based group decision strategy for event venue recommendation.
We use a widely used metric, {\bf hit rate}, to evaluate the recommendation performance of these models. 
For each group, we recommend top-N (N=5,10,15,20 in our experiments). Hit rate is defined as the proportion
of groups' top-N recommendation lists that includes the final venue decided by the group in our dataset.  For example, suppose there are 100 group events; then we would generate 100 top-N lists.  For a given event, if the final venue chosen by the participants is included in the top-N recommendation list generated for that event, then we consider that the recommendation to be a success. If out of 100 events, 30 events' final venues are included in their corresponding top-N lists, then the hit rate (final venue) is 0.3.

\begin{figure}
\includegraphics[width=0.85\linewidth]{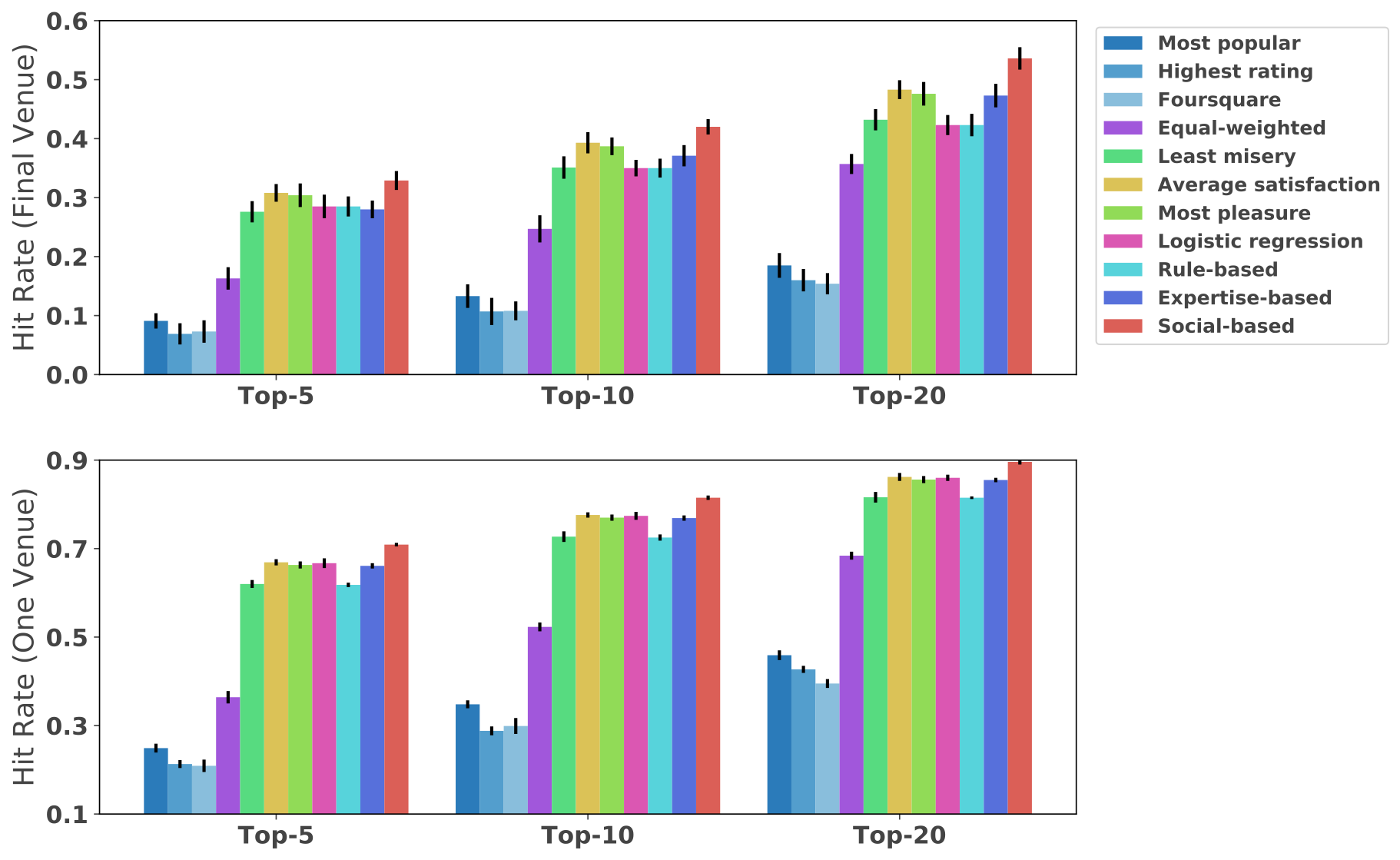}
\caption{Recommendation performance of different models on hitting a group's final venue (top subfigure) 
and hitting at least one of the group's suggested venues (bottom subfigure). 
Our \modelname{social-based} group location prediction model outperforms all the other comparative models
consistently in both cases.}
\label{fig:hitrate}
\end{figure}

Results of hitting the final group meeting venue for each of these baseline models, 
as well as our~\systemname{}, are shown in Figure~\ref{fig:hitrate}, top subfigure.
We can see that our baselines gradually build towards the performance 
of our final recommendation system
GEVR, indicating that each of the components 
(\textbf{location clusters}, \textbf{individual location preference model}, and \textbf{social-based group decision strategy}) 
is a meaningful contribution to the event venue recommendation for groups of mobile users.

The three basic baselines are clearly ineffective, 
achieving an average hit rate for Top 5, Top 10, 
and Top 20, with only 0.08, 0.12, and 0.17, respectively.
In comparison, the \modelname{equal-weighted} model, which considers group location cluster information, 
achieves better performance. 
But it is not as effective as the remaining ones that model the group decision-making process 
to help re-rank the recommendation list.
Our \modelname{social-based} group location prediction model outperforms 
the other six comparative models 
with an average improvement of 14\% for Top-5, 15\% for Top-10, 
and 19\% for Top-20.
This fits our expectation, as the social-based group decision strategy provides the most promising 
accuracy in predicting the groups' final location cluster, as shown in Table~\ref{tab:group}.

We further calculate the success rate of hitting at least
one of the group's suggested venues, i.e., venues that they voted on, using all these models. 
This is a different and broader metric than the prior hit rate (final venue), 
so we define this as hit rate (one venue).  
For example, suppose for a given event that the group members have suggested three venues 
that they are interested in visiting.  
Then if any one of those three appear in the top-N list, 
we consider that recommendation to be a success. 
If out of 100 group events, at least one of the voted-upon venues appear in a top-N list for 30 of those events, 
then the hit rate (one venue) is 0.3.  
The results of this metric are shown in Figure~\ref{fig:closest}, bottom subfigure.  
As can be seen, the relative recommendation performance of the different methods 
follows a similar pattern as the previous metric.  
The overall absolute value of this hit rate (one venue) metric is higher than the final venue hit rate 
because there is a higher chance of any one of the suggested venues appearing 
in a top-N list compared with just the final venue appearing.

\section{Concluding Discussion}
\label{sec:discussion}
In this work, we present \systemname{}, 
the first event venue recommendation system for groups of mobile users.
The system tackles the recommendation challenges by splitting the problem into three separate steps:
(1) detecting group location clusters using group members' location traces; 
(2) predicting where the group will meet; and 
(3) aggregating and re-ranking venues in nearby areas, crawled through the Foursquare Venue Recommendation API, 
to create a final venue recommendation list based on the prediction results.
When predicting a group's gathering decision, we first model individual group members' location preferences based on 
four types of features: location familiarity, user activeness, day of the week, and population density. 
We  then design a novel social-based group location prediction model to aggregate individual preferences
and predict the group's final decisions based on the group's social relationship strength.

To evaluate the performance of \systemname{}, we have collected a unique dataset from a 
group event scheduling mobile application, OutWithFriendz. 
In total, 625 users participated in this study and created 502 group events. 
Our users are widespread across the US, covering 40 states and 117 cities.
Evaluation results show that \systemname{} can provide over 80\% accuracy for predicting 
which location cluster the group will meet at, and our newly designed social-based group location prediction model 
outperforms all the other state-of-the-art group location prediction models with an average improvement of 16\% on hit rate.  

\subsection{Significant Implications} 
One of the significant implications of this work is that it is essential to integrate location considerations into an event venue recommendation framework for groups of mobile users. As demonstrated clearly in Figure~\ref{fig:hitrate}, approaches that do not consider group members' location behavior in the recommendation process performed much worse than our GEVR approach.  We expect that the research community will build upon this finding, and future solutions for group recommendation in other contexts will also incorporate mobility into their recommendation strategies.

Similarly, our work clearly demonstrates that methods for individual venue recommendation such as \modelname{most Popular,} \modelname{highest rating,} and \modelname{Foursquare} are considerably less effective when applied to group venue recommendation than algorithms that incorporate knowledge of group characteristics.  Group properties, such as mobility spread, membership fluidity, and diversity of interests, introduce a higher level of complexity that impacts the final venue chosen by the group.
Unlike individual recommendation, which mostly relies on personal preferences, 
negotiation and coordination are necessary for group members to reach a final agreement. 
Understanding the group negotiation and coordination process is critical for understanding
group event scheduling behaviors and providing effective group event venue recommendation. 
Our work also indicates that one simple group decision strategy, e.g., average satisfaction,
cannot apply to all the groups. 
Groups with strong social relationships usually apply different strategies compared with groups
with weak social relationships. 

Our results are uniquely validated based on a large number of real-world group venue events that have been quite challenging to obtain. For privacy reasons, we are unable to release the dataset of group events, because the location traces are difficult to fully anonymize.  However, we can lower the barrier to entry for future researchers and have obtained permission from the authors of the software used in this study to release the applications as open source.  This will enable other researchers to obtain valuable ground truth data to study the fine-grained behavior of groups of mobile users, and thus drive the growth in the research community for developing practically validated recommendation solutions for groups of mobile users.

Our work builds upon commercial location-based social services (LBSNs) such as Foursquare.  This strategy opens up opportunities for more in-depth partnerships with the industry on joint development of novel group recommendation algorithms and systems that can be validated in the real world.  This research strengthens the capabilities of academia in approaching the industry for collaborations, such as Google, Apple, and mobile social networks like Facebook, Instagram, and Snapchat, in hopes of introducing mobile tools that can facilitate useful group event coordination and venue recommendation.

There are a variety of strong use cases for our~\systemname{} in real life. 
First, online group event
organization services could use~\systemname{} to find venues in their databases 
that are highly likely to be suitable for most group members. 
As previous research has pointed out, for many social groups, how to organize an event
and attract more participants continues to be a difficult task. 
Even experienced organizers still feel stressed 
when planning a group event~\cite{allen2008event}.
Our~\systemname{} could to some extent reduce the group event host's 
burden of deciding where to meet. 
Additionally, although future research is needed to verify this hypothesis, 
it is possible that our~\systemname{} could help target potential participants
who would be interested in attending this group event. 
With the suggested venues and ongoing voting results of a new group event, 
we may be able to find
group members' friends who fit this proposal perfectly. 
Validating this hypothesis could also be an exciting topic for future research.

\subsection{Future Work}  
We plan to investigate the following dimensions to further improve the performance of GEVR. We would like to explore richer contextual clues in hopes of improving the accuracy of group recommendation. Location/mobility has proven to be quite helpful in improving the performance of the algorithm.  More detailed social connection information may also prove fruitful. Weather forecast information could also be leveraged in our system. In addition, more historical information, such as voting behaviors within the application and venue preferences, would be useful to explore. Our dataset is comprised primarily of younger people, and we would like to incorporate a broader age demographic among our groups.  We are also interested in how to incorporate serendipity into the group event venue recommendation process. 
Our current recommendation system is dependent on historical behavior in terms of 
preferences and locations. However, we also find that roughly 10\% of the group events occurred in venues that do not fall into any of the detected group location
clusters, indicating that groups may sometimes explore new venues that have not been visited before. 
To improve our system, we can leverage collaborative filtering, which has been shown to be very effective for individual recommendation. Specifically, when we have many groups in the same city,  we can measure the similarity of different groups based on the context information, such as location traces and venue preferences. We can then recommend venues that are frequently visited by similar groups.
Another strategy is to add a few ``most popular''
or ``highest rated'' venues in nearby areas that the group has never visited before into the recommendation list. 
How to balance the level of diversity between familiar and unfamiliar recommended places is an important research question for future exploration.

During our user study, we do not restrict the types of events that groups can organize. There are different types of events in our dataset, such as dining, outdoor hiking, and movies. However, we find that dining is the dominant event type. As described in Section~\ref{sec:data}, dining events account for 81\% of all events in our dataset. 
As such, we decided to focus on dining recommendation in this paper. This is actually more challenging than recommending other types of event venues, since the number of restaurants in the same location cluster is much higher than other types of venues, such as movie theaters or hiking trails. We expect the proposed recommendation strategy would also help for other types of events, and would like to investigate it further when we have more data for other types of events.

Privacy concerns are becoming increasingly important, and we are
interested in exploring ways to protect group members' location privacy.
Protecting members' location privacy in group events may be more challenging
than ensuring the privacy of individual users in other applications, as users'
sensitive location information may be exposed to other group members. 
There are well-known individual privacy protection mechanisms such as   
 k-anonymity and differential privacy. Whether those methods can offer adequate
privacy protection for groups in a dynamic mobile setting is an open question.
To address this problem, we would like to investigate several possible solutions.
For example, for users who have privacy concerns and do not want their locations to be tracked, we could ask them to explicitly specify a list of their
favorite venues when registering to use our application. Location preferences can be modeled using these specified venues, which may not be as effective as using location traces directly, but offers a good comprise between accuracy and privacy. 
In addition, adding a few ``most popular'' or
``highest rated'' venues in infrequently visited and unknown areas may be a
good strategy to minimize disclosure of location information among group
members.

\begin{acks}
We thank anonymous reviewers for helpful comments and discussions. 
We thank Scott Fredrick Holman from the CU Boulder Writing Center for his feedback and support during the writing process. This work is supported in part by the US National Science
Foundation (NSF) through grant CNS 1528138.
\end{acks}
\bibliographystyle{ACM-Reference-Format}
\bibliography{paper}


\begin{thebibliography}{47}


\ifx \showCODEN    \undefined \def \showCODEN     #1{\unskip}     \fi
\ifx \showDOI      \undefined \def \showDOI       #1{#1}\fi
\ifx \showISBNx    \undefined \def \showISBNx     #1{\unskip}     \fi
\ifx \showISBNxiii \undefined \def \showISBNxiii  #1{\unskip}     \fi
\ifx \showISSN     \undefined \def \showISSN      #1{\unskip}     \fi
\ifx \showLCCN     \undefined \def \showLCCN      #1{\unskip}     \fi
\ifx \shownote     \undefined \def \shownote      #1{#1}          \fi
\ifx \showarticletitle \undefined \def \showarticletitle #1{#1}   \fi
\ifx \showURL      \undefined \def \showURL       {\relax}        \fi
\providecommand\bibfield[2]{#2}
\providecommand\bibinfo[2]{#2}
\providecommand\natexlab[1]{#1}
\providecommand\showeprint[2][]{arXiv:#2}

\bibitem[\protect\citeauthoryear{Agatz, Erera, Savelsbergh, and Wang}{Agatz
  et~al\mbox{.}}{2012}]%
        {agatz2012optimization}
\bibfield{author}{\bibinfo{person}{Niels Agatz}, \bibinfo{person}{Alan Erera},
  \bibinfo{person}{Martin Savelsbergh}, {and} \bibinfo{person}{Xing Wang}.}
  \bibinfo{year}{2012}\natexlab{}.
\newblock \showarticletitle{Optimization for dynamic ride-sharing: A review}.
\newblock \bibinfo{journal}{\emph{European Journal of Operational Research}}
  \bibinfo{volume}{223}, \bibinfo{number}{2} (\bibinfo{year}{2012}),
  \bibinfo{pages}{295--303}.
\newblock


\bibitem[\protect\citeauthoryear{Allen}{Allen}{2008}]%
        {allen2008event}
\bibfield{author}{\bibinfo{person}{Judy Allen}.}
  \bibinfo{year}{2008}\natexlab{}.
\newblock \bibinfo{booktitle}{\emph{Event planning: The ultimate guide to
  successful meetings, corporate events, fundraising galas, conferences,
  conventions, incentives and other special events}}.
\newblock \bibinfo{publisher}{John Wiley \& Sons}.
\newblock


\bibitem[\protect\citeauthoryear{Ashbrook and Starner}{Ashbrook and
  Starner}{2003}]%
        {ashbrook2003using}
\bibfield{author}{\bibinfo{person}{Daniel Ashbrook} {and} \bibinfo{person}{Thad
  Starner}.} \bibinfo{year}{2003}\natexlab{}.
\newblock \showarticletitle{Using GPS to learn significant locations and
  predict movement across multiple users}.
\newblock \bibinfo{journal}{\emph{Personal and Ubiquitous Computing}}
  \bibinfo{volume}{7}, \bibinfo{number}{5} (\bibinfo{year}{2003}),
  \bibinfo{pages}{275--286}.
\newblock


\bibitem[\protect\citeauthoryear{Backstrom, Sun, and Marlow}{Backstrom
  et~al\mbox{.}}{2010}]%
        {backstrom2010find}
\bibfield{author}{\bibinfo{person}{Lars Backstrom}, \bibinfo{person}{Eric Sun},
  {and} \bibinfo{person}{Cameron Marlow}.} \bibinfo{year}{2010}\natexlab{}.
\newblock \showarticletitle{Find me if you can: improving geographical
  prediction with social and spatial proximity}. In
  \bibinfo{booktitle}{\emph{Proceedings of WWW}}. \bibinfo{pages}{61--70}.
\newblock


\bibitem[\protect\citeauthoryear{Baumann, Kleiminger, and Santini}{Baumann
  et~al\mbox{.}}{2013}]%
        {baumann2013influence}
\bibfield{author}{\bibinfo{person}{Paul Baumann}, \bibinfo{person}{Wilhelm
  Kleiminger}, {and} \bibinfo{person}{Silvia Santini}.}
  \bibinfo{year}{2013}\natexlab{}.
\newblock \showarticletitle{The influence of temporal and spatial features on
  the performance of next-place prediction algorithms}. In
  \bibinfo{booktitle}{\emph{Proceedings of UbiComp}}.
  \bibinfo{pages}{449--458}.
\newblock


\bibitem[\protect\citeauthoryear{Beckmann and Gross}{Beckmann and
  Gross}{2011}]%
        {beckmann2011agremo}
\bibfield{author}{\bibinfo{person}{Christoph Beckmann} {and}
  \bibinfo{person}{Tom Gross}.} \bibinfo{year}{2011}\natexlab{}.
\newblock \showarticletitle{AGReMo: providing ad-hoc groups with on-demand
  recommendations on mobile devices}. In \bibinfo{booktitle}{\emph{Proceedings
  of the 29th Annual European Conference on Cognitive Ergonomics}}. ACM,
  \bibinfo{pages}{179--182}.
\newblock


\bibitem[\protect\citeauthoryear{Bureau}{Bureau}{2016}]%
        {uscensus}
\bibfield{author}{\bibinfo{person}{US~Census Bureau}.}
  \bibinfo{year}{2016}\natexlab{}.
\newblock \bibinfo{title}{US Population Density Dataset 2010-2016}.
\newblock
  \bibinfo{howpublished}{\url{https://www2.census.gov/programs-surveys/popest/datasets/2010-2016/counties/totals/co-est2016-alldata.csv}}.
\newblock
\newblock
\shownote{[Online; accessed 01-November-2018].}


\bibitem[\protect\citeauthoryear{Chaney, Gartrell, Hofman, Guiver, Koenigstein,
  Kohli, and Paquet}{Chaney et~al\mbox{.}}{2014}]%
        {chaney2014large}
\bibfield{author}{\bibinfo{person}{Allison~JB Chaney}, \bibinfo{person}{Mike
  Gartrell}, \bibinfo{person}{Jake~M Hofman}, \bibinfo{person}{John Guiver},
  \bibinfo{person}{Noam Koenigstein}, \bibinfo{person}{Pushmeet Kohli}, {and}
  \bibinfo{person}{Ulrich Paquet}.} \bibinfo{year}{2014}\natexlab{}.
\newblock \showarticletitle{A large-scale exploration of group viewing
  patterns}. In \bibinfo{booktitle}{\emph{Proceedings of TVX}}.
  \bibinfo{pages}{31--38}.
\newblock


\bibitem[\protect\citeauthoryear{Chen, Zhang, Wang, Yang, Ma, Li, Wu, Pan,
  Nguyen, and Jakubowicz}{Chen et~al\mbox{.}}{2016}]%
        {chen2016dynamic}
\bibfield{author}{\bibinfo{person}{Longbiao Chen}, \bibinfo{person}{Daqing
  Zhang}, \bibinfo{person}{Leye Wang}, \bibinfo{person}{Dingqi Yang},
  \bibinfo{person}{Xiaojuan Ma}, \bibinfo{person}{Shijian Li},
  \bibinfo{person}{Zhaohui Wu}, \bibinfo{person}{Gang Pan},
  \bibinfo{person}{Thi-Mai-Trang Nguyen}, {and}
  \bibinfo{person}{J{\'e}r{\'e}mie Jakubowicz}.}
  \bibinfo{year}{2016}\natexlab{}.
\newblock \showarticletitle{Dynamic cluster-based over-demand prediction in
  bike sharing systems}. In \bibinfo{booktitle}{\emph{Proceedings of UbiComp}}.
  \bibinfo{pages}{841--852}.
\newblock


\bibitem[\protect\citeauthoryear{Cici, Markopoulou, Frias-Martinez, and
  Laoutaris}{Cici et~al\mbox{.}}{2014}]%
        {cici2014assessing}
\bibfield{author}{\bibinfo{person}{Blerim Cici}, \bibinfo{person}{Athina
  Markopoulou}, \bibinfo{person}{Enrique Frias-Martinez}, {and}
  \bibinfo{person}{Nikolaos Laoutaris}.} \bibinfo{year}{2014}\natexlab{}.
\newblock \showarticletitle{Assessing the potential of ride-sharing using
  mobile and social data: a tale of four cities}. In
  \bibinfo{booktitle}{\emph{Proceedings of the 2014 ACM International Joint
  Conference on Pervasive and Ubiquitous Computing}}. ACM,
  \bibinfo{pages}{201--211}.
\newblock


\bibitem[\protect\citeauthoryear{Du, Yu, Mei, Wang, Wang, and Guo}{Du
  et~al\mbox{.}}{2014}]%
        {du2014predicting}
\bibfield{author}{\bibinfo{person}{Rong Du}, \bibinfo{person}{Zhiwen Yu},
  \bibinfo{person}{Tao Mei}, \bibinfo{person}{Zhitao Wang},
  \bibinfo{person}{Zhu Wang}, {and} \bibinfo{person}{Bin Guo}.}
  \bibinfo{year}{2014}\natexlab{}.
\newblock \showarticletitle{Predicting activity attendance in event-based
  social networks: Content, context and social influence}. In
  \bibinfo{booktitle}{\emph{Proceedings of UbiComp}}. ACM,
  \bibinfo{pages}{425--434}.
\newblock


\bibitem[\protect\citeauthoryear{Elgar and Bekhor}{Elgar and Bekhor}{2004}]%
        {elgar2004car}
\bibfield{author}{\bibinfo{person}{Alon Elgar} {and} \bibinfo{person}{Shlomo
  Bekhor}.} \bibinfo{year}{2004}\natexlab{}.
\newblock \showarticletitle{Car-rider segmentation according to riding status
  and investment in car mobility}.
\newblock \bibinfo{journal}{\emph{Transportation Research Record: Journal of
  the Transportation Research Board}} \bibinfo{number}{1894}
  (\bibinfo{year}{2004}), \bibinfo{pages}{109--116}.
\newblock


\bibitem[\protect\citeauthoryear{Ester, Kriegel, Sander, and Xu}{Ester
  et~al\mbox{.}}{1996}]%
        {ester1996density}
\bibfield{author}{\bibinfo{person}{Martin Ester}, \bibinfo{person}{Hans-Peter
  Kriegel}, \bibinfo{person}{J\"{o}rg Sander}, {and} \bibinfo{person}{Xiaowei
  Xu}.} \bibinfo{year}{1996}\natexlab{}.
\newblock \showarticletitle{A Density-based Algorithm for Discovering Clusters
  a Density-based Algorithm for Discovering Clusters in Large Spatial Databases
  with Noise}. In \bibinfo{booktitle}{\emph{Proceedings of KDD}}.
\newblock


\bibitem[\protect\citeauthoryear{Foursquare}{Foursquare}{2018}]%
        {foursquare}
\bibfield{author}{\bibinfo{person}{Foursquare}.}
  \bibinfo{year}{2018}\natexlab{}.
\newblock \bibinfo{title}{Foursquare Venue Recomendation API}.
\newblock
  \bibinfo{howpublished}{\url{https://developer.foursquare.com/docs/api/venues/explore}}.
\newblock
\newblock
\shownote{[Online; accessed 01-November-2018].}


\bibitem[\protect\citeauthoryear{Gartrell, Xing, Lv, Beach, Han, Mishra, and
  Seada}{Gartrell et~al\mbox{.}}{2010}]%
        {gartrell2010enhancing}
\bibfield{author}{\bibinfo{person}{Mike Gartrell}, \bibinfo{person}{Xinyu
  Xing}, \bibinfo{person}{Qin Lv}, \bibinfo{person}{Aaron Beach},
  \bibinfo{person}{Richard Han}, \bibinfo{person}{Shivakant Mishra}, {and}
  \bibinfo{person}{Karim Seada}.} \bibinfo{year}{2010}\natexlab{}.
\newblock \showarticletitle{Enhancing group recommendation by incorporating
  social relationship interactions}. In \bibinfo{booktitle}{\emph{Proceedings
  of ACM Group}}. \bibinfo{pages}{97--106}.
\newblock


\bibitem[\protect\citeauthoryear{Georgiev, Noulas, and Mascolo}{Georgiev
  et~al\mbox{.}}{2014}]%
        {georgiev2014call}
\bibfield{author}{\bibinfo{person}{Petko~Ivanov Georgiev},
  \bibinfo{person}{Anastasios Noulas}, {and} \bibinfo{person}{Cecilia
  Mascolo}.} \bibinfo{year}{2014}\natexlab{}.
\newblock \showarticletitle{The Call of the Crowd: Event Participation in
  Location-Based Social Services}. In \bibinfo{booktitle}{\emph{Proceedings of
  ICWSM}}. \bibinfo{pages}{141--150}.
\newblock


\bibitem[\protect\citeauthoryear{Gilbert, Karahalios, and Sandvig}{Gilbert
  et~al\mbox{.}}{2008}]%
        {gilbert2008network}
\bibfield{author}{\bibinfo{person}{Eric Gilbert}, \bibinfo{person}{Karrie
  Karahalios}, {and} \bibinfo{person}{Christian Sandvig}.}
  \bibinfo{year}{2008}\natexlab{}.
\newblock \showarticletitle{The network in the garden: an empirical analysis of
  social media in rural life}. In \bibinfo{booktitle}{\emph{Proceedings of
  CHI}}. \bibinfo{pages}{1603--1612}.
\newblock


\bibitem[\protect\citeauthoryear{Hemminki, Nurmi, and Tarkoma}{Hemminki
  et~al\mbox{.}}{2013}]%
        {hemminki2013accelerometer}
\bibfield{author}{\bibinfo{person}{Samuli Hemminki}, \bibinfo{person}{Petteri
  Nurmi}, {and} \bibinfo{person}{Sasu Tarkoma}.}
  \bibinfo{year}{2013}\natexlab{}.
\newblock \showarticletitle{Accelerometer-based transportation mode detection
  on smartphones}. In \bibinfo{booktitle}{\emph{Proceedings of SenSys}}.
  \bibinfo{pages}{13}.
\newblock


\bibitem[\protect\citeauthoryear{Hsieh, Tangmunarunkit, Alquaddoomi, Jenkins,
  Kang, Ketcham, Longstaff, Selsky, Dawson, Swendeman, et~al\mbox{.}}{Hsieh
  et~al\mbox{.}}{2013}]%
        {hsieh2013lifestreams}
\bibfield{author}{\bibinfo{person}{Cheng-Kang Hsieh}, \bibinfo{person}{Hongsuda
  Tangmunarunkit}, \bibinfo{person}{Faisal Alquaddoomi}, \bibinfo{person}{John
  Jenkins}, \bibinfo{person}{Jinha Kang}, \bibinfo{person}{Cameron Ketcham},
  \bibinfo{person}{Brent Longstaff}, \bibinfo{person}{Joshua Selsky},
  \bibinfo{person}{Betta Dawson}, \bibinfo{person}{Dallas Swendeman},
  {et~al\mbox{.}}} \bibinfo{year}{2013}\natexlab{}.
\newblock \showarticletitle{Lifestreams: A modular sense-making toolset for
  identifying important patterns from everyday life}. In
  \bibinfo{booktitle}{\emph{Proceedings of SenSys}}.
\newblock


\bibitem[\protect\citeauthoryear{Jameson and Smyth}{Jameson and Smyth}{2007}]%
        {jameson2007recommendation}
\bibfield{author}{\bibinfo{person}{Anthony Jameson} {and}
  \bibinfo{person}{Barry Smyth}.} \bibinfo{year}{2007}\natexlab{}.
\newblock \showarticletitle{Recommendation to groups}.
\newblock In \bibinfo{booktitle}{\emph{The Adaptive Web}}.
  \bibinfo{publisher}{Springer}, \bibinfo{pages}{596--627}.
\newblock


\bibitem[\protect\citeauthoryear{Jayarajah, Lee, Misra, and Balan}{Jayarajah
  et~al\mbox{.}}{2015}]%
        {jayarajah2015need}
\bibfield{author}{\bibinfo{person}{Kasthuri Jayarajah},
  \bibinfo{person}{Youngki Lee}, \bibinfo{person}{Archan Misra}, {and}
  \bibinfo{person}{Rajesh~Krishna Balan}.} \bibinfo{year}{2015}\natexlab{}.
\newblock \showarticletitle{Need accurate user behaviour?: pay attention to
  groups!}. In \bibinfo{booktitle}{\emph{Proceedings of UbiComp}}. ACM,
  \bibinfo{pages}{855--866}.
\newblock


\bibitem[\protect\citeauthoryear{Lane, Pengyu, Zhou, and Zhao}{Lane
  et~al\mbox{.}}{2014}]%
        {lane2014connecting}
\bibfield{author}{\bibinfo{person}{Nicholas~D Lane}, \bibinfo{person}{Li
  Pengyu}, \bibinfo{person}{Lin Zhou}, {and} \bibinfo{person}{Feng Zhao}.}
  \bibinfo{year}{2014}\natexlab{}.
\newblock \showarticletitle{Connecting personal-scale sensing and networked
  community behavior to infer human activities}. In
  \bibinfo{booktitle}{\emph{Proceedings of the 2014 ACM International Joint
  Conference on Pervasive and Ubiquitous Computing}}. ACM,
  \bibinfo{pages}{595--606}.
\newblock


\bibitem[\protect\citeauthoryear{Lian, Zhao, Xie, Sun, Chen, and Rui}{Lian
  et~al\mbox{.}}{2014}]%
        {lian2014geomf}
\bibfield{author}{\bibinfo{person}{Defu Lian}, \bibinfo{person}{Cong Zhao},
  \bibinfo{person}{Xing Xie}, \bibinfo{person}{Guangzhong Sun},
  \bibinfo{person}{Enhong Chen}, {and} \bibinfo{person}{Yong Rui}.}
  \bibinfo{year}{2014}\natexlab{}.
\newblock \showarticletitle{GeoMF: joint geographical modeling and matrix
  factorization for point-of-interest recommendation}. In
  \bibinfo{booktitle}{\emph{Proceedings of KDD}}. \bibinfo{pages}{831--840}.
\newblock


\bibitem[\protect\citeauthoryear{Loh}{Loh}{2011}]%
        {loh2011classification}
\bibfield{author}{\bibinfo{person}{Wei-Yin Loh}.}
  \bibinfo{year}{2011}\natexlab{}.
\newblock \showarticletitle{Classification and regression trees}.
\newblock \bibinfo{journal}{\emph{Wiley Interdisciplinary Reviews: Data Mining
  and Knowledge Discovery}} \bibinfo{volume}{1}, \bibinfo{number}{1}
  (\bibinfo{year}{2011}), \bibinfo{pages}{14--23}.
\newblock


\bibitem[\protect\citeauthoryear{Lu, Wang, Yang, Pang, and Zhang}{Lu
  et~al\mbox{.}}{2010}]%
        {lu2010photo2trip}
\bibfield{author}{\bibinfo{person}{Xin Lu}, \bibinfo{person}{Changhu Wang},
  \bibinfo{person}{Jiang-Ming Yang}, \bibinfo{person}{Yanwei Pang}, {and}
  \bibinfo{person}{Lei Zhang}.} \bibinfo{year}{2010}\natexlab{}.
\newblock \showarticletitle{Photo2trip: generating travel routes from
  geo-tagged photos for trip planning}. In
  \bibinfo{booktitle}{\emph{Proceedings of ICMI}}. \bibinfo{pages}{143--152}.
\newblock


\bibitem[\protect\citeauthoryear{Macedo, Marinho, and Santos}{Macedo
  et~al\mbox{.}}{2015}]%
        {macedo2015context}
\bibfield{author}{\bibinfo{person}{Augusto~Q Macedo},
  \bibinfo{person}{Leandro~B Marinho}, {and} \bibinfo{person}{Rodrygo~LT
  Santos}.} \bibinfo{year}{2015}\natexlab{}.
\newblock \showarticletitle{Context-aware event recommendation in event-based
  social networks}. In \bibinfo{booktitle}{\emph{Proceedings of RecSys}}.
  \bibinfo{pages}{123--130}.
\newblock


\bibitem[\protect\citeauthoryear{Masthoff}{Masthoff}{2004}]%
        {masthoff2004group}
\bibfield{author}{\bibinfo{person}{Judith Masthoff}.}
  \bibinfo{year}{2004}\natexlab{}.
\newblock \showarticletitle{Group modeling: Selecting a sequence of television
  items to suit a group of viewers}. In \bibinfo{booktitle}{\emph{Personalized
  digital television}}. \bibinfo{publisher}{Springer},
  \bibinfo{pages}{93--141}.
\newblock


\bibitem[\protect\citeauthoryear{Noulas, Scellato, Lathia, and Mascolo}{Noulas
  et~al\mbox{.}}{2012a}]%
        {noulas2012mining}
\bibfield{author}{\bibinfo{person}{Anastasios Noulas},
  \bibinfo{person}{Salvatore Scellato}, \bibinfo{person}{Neal Lathia}, {and}
  \bibinfo{person}{Cecilia Mascolo}.} \bibinfo{year}{2012}\natexlab{a}.
\newblock \showarticletitle{Mining user mobility features for next place
  prediction in location-based services}. In
  \bibinfo{booktitle}{\emph{Proceedings of ICDM}}. \bibinfo{pages}{1038--1043}.
\newblock


\bibitem[\protect\citeauthoryear{Noulas, Scellato, Lathia, and Mascolo}{Noulas
  et~al\mbox{.}}{2012b}]%
        {noulas2012random}
\bibfield{author}{\bibinfo{person}{Anastasios Noulas},
  \bibinfo{person}{Salvatore Scellato}, \bibinfo{person}{Neal Lathia}, {and}
  \bibinfo{person}{Cecilia Mascolo}.} \bibinfo{year}{2012}\natexlab{b}.
\newblock \showarticletitle{A random walk around the city: New venue
  recommendation in location-based social networks}. In
  \bibinfo{booktitle}{\emph{Privacy, Security, Risk and Trust (PASSAT), 2012
  International Conference on Social Computing}}. IEEE,
  \bibinfo{pages}{144--153}.
\newblock


\bibitem[\protect\citeauthoryear{Park, Lim, Kim, Lee, and Lee}{Park
  et~al\mbox{.}}{2017}]%
        {park2017don}
\bibfield{author}{\bibinfo{person}{Chunjong Park}, \bibinfo{person}{Junsung
  Lim}, \bibinfo{person}{Juho Kim}, \bibinfo{person}{Sung-Ju Lee}, {and}
  \bibinfo{person}{Dongman Lee}.} \bibinfo{year}{2017}\natexlab{}.
\newblock \showarticletitle{{Don't Bother Me. I'm Socializing!: A
  Breakpoint-Based Smartphone Notification System}}. In
  \bibinfo{booktitle}{\emph{Proceedings of CSCW}}. \bibinfo{pages}{541--554}.
\newblock


\bibitem[\protect\citeauthoryear{Park, Park, and Cho}{Park
  et~al\mbox{.}}{2008}]%
        {park2008restaurant}
\bibfield{author}{\bibinfo{person}{Moon-Hee Park}, \bibinfo{person}{Han-Saem
  Park}, {and} \bibinfo{person}{Sung-Bae Cho}.}
  \bibinfo{year}{2008}\natexlab{}.
\newblock \showarticletitle{Restaurant recommendation for group of people in
  mobile environments using probabilistic multi-criteria decision making}. In
  \bibinfo{booktitle}{\emph{Asia-Pacific Conference on Computer Human
  Interaction}}. Springer, \bibinfo{pages}{114--122}.
\newblock


\bibitem[\protect\citeauthoryear{Quintarelli, Rabosio, and Tanca}{Quintarelli
  et~al\mbox{.}}{2016}]%
        {quintarelli2016recommending}
\bibfield{author}{\bibinfo{person}{Elisa Quintarelli},
  \bibinfo{person}{Emanuele Rabosio}, {and} \bibinfo{person}{Letizia Tanca}.}
  \bibinfo{year}{2016}\natexlab{}.
\newblock \showarticletitle{Recommending new items to ephemeral groups using
  contextual user influence}. In \bibinfo{booktitle}{\emph{Proceedings of
  RecSys}}. \bibinfo{pages}{285--292}.
\newblock


\bibitem[\protect\citeauthoryear{Ramos et~al\mbox{.}}{Ramos
  et~al\mbox{.}}{2003}]%
        {ramos2003using}
\bibfield{author}{\bibinfo{person}{Juan Ramos} {et~al\mbox{.}}}
  \bibinfo{year}{2003}\natexlab{}.
\newblock \showarticletitle{Using tf-idf to determine word relevance in
  document queries}. In \bibinfo{booktitle}{\emph{Proceedings of the first
  instructional conference on machine learning}}, Vol.~\bibinfo{volume}{242}.
  \bibinfo{pages}{133--142}.
\newblock


\bibitem[\protect\citeauthoryear{Reinecke, Nguyen, Bernstein, N{\"a}f, and
  Gajos}{Reinecke et~al\mbox{.}}{2013}]%
        {reinecke2013doodle}
\bibfield{author}{\bibinfo{person}{Katharina Reinecke},
  \bibinfo{person}{Minh~Khoa Nguyen}, \bibinfo{person}{Abraham Bernstein},
  \bibinfo{person}{Michael N{\"a}f}, {and} \bibinfo{person}{Krzysztof~Z
  Gajos}.} \bibinfo{year}{2013}\natexlab{}.
\newblock \showarticletitle{Doodle around the world: Online scheduling behavior
  reflects cultural differences in time perception and group decision-making}.
  In \bibinfo{booktitle}{\emph{Proceedings of CSCW}}. \bibinfo{pages}{45--54}.
\newblock


\bibitem[\protect\citeauthoryear{Romero, Reinecke, and Robert~Jr}{Romero
  et~al\mbox{.}}{2017}]%
        {romero2017influence}
\bibfield{author}{\bibinfo{person}{Daniel~M Romero}, \bibinfo{person}{Katharina
  Reinecke}, {and} \bibinfo{person}{Lionel~P Robert~Jr}.}
  \bibinfo{year}{2017}\natexlab{}.
\newblock \showarticletitle{The Influence of Early Respondents: Information
  Cascade Effects in Online Event Scheduling}. In
  \bibinfo{booktitle}{\emph{Proceedings of WSDM}}. \bibinfo{pages}{101--110}.
\newblock


\bibitem[\protect\citeauthoryear{Source}{Source}{2017}]%
        {weathersource}
\bibfield{author}{\bibinfo{person}{Weather Source}.}
  \bibinfo{year}{2017}\natexlab{}.
\newblock \bibinfo{title}{Weather Source API}.
\newblock
  \bibinfo{howpublished}{\url{https://weathersource.com/products/onpoint-api/}}.
\newblock
\newblock
\shownote{[Online; accessed 01-November-2018].}


\bibitem[\protect\citeauthoryear{Stenneth, Wolfson, Yu, and Xu}{Stenneth
  et~al\mbox{.}}{2011}]%
        {stenneth2011transportation}
\bibfield{author}{\bibinfo{person}{Leon Stenneth}, \bibinfo{person}{Ouri
  Wolfson}, \bibinfo{person}{Philip~S Yu}, {and} \bibinfo{person}{Bo Xu}.}
  \bibinfo{year}{2011}\natexlab{}.
\newblock \showarticletitle{Transportation mode detection using mobile phones
  and GIS information}. In \bibinfo{booktitle}{\emph{Proceedings of
  SIGSPATIAL}}. \bibinfo{pages}{54--63}.
\newblock


\bibitem[\protect\citeauthoryear{Tong, Faloutsos, and Pan}{Tong
  et~al\mbox{.}}{2006}]%
        {tong2006fast}
\bibfield{author}{\bibinfo{person}{Hanghang Tong}, \bibinfo{person}{Christos
  Faloutsos}, {and} \bibinfo{person}{Jia-yu Pan}.}
  \bibinfo{year}{2006}\natexlab{}.
\newblock \showarticletitle{Fast Random Walk with Restart and Its
  Applications}. In \bibinfo{booktitle}{\emph{Proceedings of ICDM}}.
  \bibinfo{pages}{613--622}.
\newblock


\bibitem[\protect\citeauthoryear{Wikipedia}{Wikipedia}{2018a}]%
        {craigslist}
\bibfield{author}{\bibinfo{person}{Wikipedia}.}
  \bibinfo{year}{2018}\natexlab{a}.
\newblock \bibinfo{title}{{Craigslist} --- {W}ikipedia{,} The Free
  Encyclopedia}.
\newblock
  \bibinfo{howpublished}{\url{http://en.wikipedia.org/w/index.php?title=Craigslist&oldid=836044084}}.
\newblock
\newblock
\shownote{[Online; accessed 01-November-2018].}


\bibitem[\protect\citeauthoryear{Wikipedia}{Wikipedia}{2018b}]%
        {microworkers}
\bibfield{author}{\bibinfo{person}{Wikipedia}.}
  \bibinfo{year}{2018}\natexlab{b}.
\newblock \bibinfo{title}{{Microwork} --- {W}ikipedia{,} The Free
  Encyclopedia}.
\newblock
  \bibinfo{howpublished}{\url{http://en.wikipedia.org/w/index.php?title=Microwork&oldid=828297165}}.
\newblock
\newblock
\shownote{[Online; accessed 01-November-2018].}


\bibitem[\protect\citeauthoryear{Zhang, Alanezi, Gartrell, Han, Lv, and
  Mishra}{Zhang et~al\mbox{.}}{2018a}]%
        {zhang2018understanding}
\bibfield{author}{\bibinfo{person}{Jason~Shuo Zhang}, \bibinfo{person}{Khaled
  Alanezi}, \bibinfo{person}{Mike Gartrell}, \bibinfo{person}{Richard Han},
  \bibinfo{person}{Qin Lv}, {and} \bibinfo{person}{Shivakant Mishra}.}
  \bibinfo{year}{2018}\natexlab{a}.
\newblock \showarticletitle{Understanding Group Event Scheduling via the
  OutWithFriendz Mobile Application}.
\newblock \bibinfo{journal}{\emph{Proceedings of the ACM on Interactive,
  Mobile, Wearable and Ubiquitous Technologies}} \bibinfo{volume}{1},
  \bibinfo{number}{4} (\bibinfo{year}{2018}), \bibinfo{pages}{175}.
\newblock


\bibitem[\protect\citeauthoryear{Zhang and Lv}{Zhang and Lv}{2018}]%
        {zhang2018hybrid}
\bibfield{author}{\bibinfo{person}{Jason~Shuo Zhang} {and} \bibinfo{person}{Qin
  Lv}.} \bibinfo{year}{2018}\natexlab{}.
\newblock \showarticletitle{Hybrid EGU-based group event participation
  prediction in event-based social networks}.
\newblock \bibinfo{journal}{\emph{Knowledge-Based Systems}}
  \bibinfo{volume}{143} (\bibinfo{year}{2018}), \bibinfo{pages}{19--29}.
\newblock


\bibitem[\protect\citeauthoryear{Zhang and Lv}{Zhang and Lv}{2019}]%
        {Zhang:2019:UEO:3306498.3243227}
\bibfield{author}{\bibinfo{person}{Jason~Shuo Zhang} {and} \bibinfo{person}{Qin
  Lv}.} \bibinfo{year}{2019}\natexlab{}.
\newblock \showarticletitle{Understanding Event Organization at Scale in
  Event-Based Social Networks}.
\newblock \bibinfo{journal}{\emph{ACM Trans. Intell. Syst. Technol.}}
  \bibinfo{volume}{10}, \bibinfo{number}{2}, Article \bibinfo{articleno}{16}
  (\bibinfo{date}{Jan.} \bibinfo{year}{2019}), \bibinfo{numpages}{23}~pages.
\newblock
\showISSN{2157-6904}


\bibitem[\protect\citeauthoryear{Zhang, Tan, and Lv}{Zhang
  et~al\mbox{.}}{2018b}]%
        {zhang2018we}
\bibfield{author}{\bibinfo{person}{Jason~Shuo Zhang}, \bibinfo{person}{Chenhao
  Tan}, {and} \bibinfo{person}{Qin Lv}.} \bibinfo{year}{2018}\natexlab{b}.
\newblock \showarticletitle{This is why we play: Characterizing Online Fan
  Communities of the NBA Teams}.
\newblock \bibinfo{journal}{\emph{Proceedings of the ACM on Human-Computer
  Interaction}} \bibinfo{volume}{2}, \bibinfo{number}{CSCW}
  (\bibinfo{year}{2018}), \bibinfo{pages}{197}.
\newblock


\bibitem[\protect\citeauthoryear{Zheng, Liu, Wang, and Xie}{Zheng
  et~al\mbox{.}}{2008}]%
        {zheng2008learning}
\bibfield{author}{\bibinfo{person}{Yu Zheng}, \bibinfo{person}{Like Liu},
  \bibinfo{person}{Longhao Wang}, {and} \bibinfo{person}{Xing Xie}.}
  \bibinfo{year}{2008}\natexlab{}.
\newblock \showarticletitle{Learning transportation mode from raw GPS data for
  geographic applications on the web}. In \bibinfo{booktitle}{\emph{Proceedings
  of WWW}}. \bibinfo{pages}{247--256}.
\newblock


\bibitem[\protect\citeauthoryear{Zheng, Zhang, Xie, and Ma}{Zheng
  et~al\mbox{.}}{2009}]%
        {zheng2009mining}
\bibfield{author}{\bibinfo{person}{Yu Zheng}, \bibinfo{person}{Lizhu Zhang},
  \bibinfo{person}{Xing Xie}, {and} \bibinfo{person}{Wei-Ying Ma}.}
  \bibinfo{year}{2009}\natexlab{}.
\newblock \showarticletitle{Mining interesting locations and travel sequences
  from GPS trajectories}. In \bibinfo{booktitle}{\emph{Proceedings of WWW}}.
  \bibinfo{pages}{791--800}.
\newblock


\bibitem[\protect\citeauthoryear{Zou, Meir, and Parkes}{Zou
  et~al\mbox{.}}{2015}]%
        {zou2015strategic}
\bibfield{author}{\bibinfo{person}{James Zou}, \bibinfo{person}{Reshef Meir},
  {and} \bibinfo{person}{David Parkes}.} \bibinfo{year}{2015}\natexlab{}.
\newblock \showarticletitle{Strategic voting behavior in doodle polls}. In
  \bibinfo{booktitle}{\emph{Proceedings CSCW}}. \bibinfo{pages}{464--472}.
\newblock


\end{thebibliography}

\end{document}